\documentclass[12pt,preprint]{aastex}

\newcommand{\etal}{et al. }
\newcommand{\myemail}{szgyula@mpe.mpg.de}
\newcommand{\cm}{{\rm cm}}
\newcommand{\erg}{{\rm erg}}
\newcommand{\s}{{\rm s}}
\newcommand{\keV}{{\rm keV}}
\newcommand\newa{{New~Astronomy}}
\newcommand\mess{{The~Messenger}}
\newcommand\sinn{{\rm sinn}}

\shorttitle{CDFS spectroscopic follow-up}
\shortauthors{Szokoly et al.}

\begin{document}

\title{The Chandra Deep Field South: \\
  Optical Spectroscopy I.
\footnote{Based on observations collected at the European
 Southern Observatory, Chile (ESO N$^{\rm o}$ 66.A-0270(A) and 67.A-0418(A)).}
}

\author{G. P. Szokoly\altaffilmark{1,2},
	J. Bergeron\altaffilmark{3},	
	G. Hasinger\altaffilmark{1,2},	
	I. Lehmann\altaffilmark{1}.	
	L. Kewley\altaffilmark{4,5},	
	V. Mainieri\altaffilmark{6},
	M. Nonino\altaffilmark{7},
	P. Rosati\altaffilmark{6},	
	R. Giacconi\altaffilmark{8},
	R. Gilli\altaffilmark{9},
	R. Gilmozzi\altaffilmark{6},
	C. Norman\altaffilmark{4},
	M. Romaniello\altaffilmark{6}
	E. Schreier\altaffilmark{8,10},
	P. Tozzi\altaffilmark{7},
	J. X. Wang\altaffilmark{4},
	W. Zheng\altaffilmark{4}
	and
	A. Zirm\altaffilmark{4}}
\email{\myemail}

\altaffiltext{1}{Max-Planck-Institut f\"ur extraterrestrische Physik, 
Giessenbachstra{\ss}e, Garching, D-85748 Germany}

\altaffiltext{2}{Astrophysikalisches Institute Potsdam, An der
Sternwarte 16, Potsdam, D-14482, Germany}

\altaffiltext{3}{Institut d'Astrophysique de Paris, 98bis, bd Arago,
75014 Paris, France}

\altaffiltext{4}{The Johns Hopkins University, Department of Physics and
Astronomy, Baltimore, MD 21218, USA}

\altaffiltext{5}{Harvard-Smithsonian Center for Astrophysics, 60
Garden Street, Cambridge, MA 02138, USA}

\altaffiltext{6}{European Southern Observatory,
Karl-Schwarzschild-Strasse 2, Garching, D-85748, Germany}

\altaffiltext{7}{Osservatorio Astronomico, Via G. Tiepolo 11, 34131
Trieste, Italy}

\altaffiltext{8}{Associated Universities, Inc. 1400 16th Stret, NW,
Suite 730, Washington, DC 20036, USA}

\altaffiltext{9}{Osservatorio Astrofisico di Arcetri, Largo E. Fermi 5,
I-50125 Firenze, Italy}

\altaffiltext{10}{Space Telescope Science
Institute, 3700 San Martin Drive, Baltimore, MD 21218, USA}

\begin{abstract}
We present the results of our spectroscopic follow-up program 
of the X-ray sources detected in the 942 ks exposure
of the {\em C}handra {\em D}eep {\em F}ield {\em S}outh (CDFS).
288 possible counterparts were observed at the VLT with the FORS1/FORS2
spectrographs for 251 
of the 349 Chandra sources
(including three additional faint X-ray sources). Spectra and R-band images are 
shown for all the observed sources and R$-$K colours are given for 
most of them. Spectroscopic redshifts were obtained for 168 
X-ray sources, 
of which 137 have both reliable optical identification and 
redshift estimate (including 16 external identifications). The R$ < 24$
observed sample comprises 161 X-ray objects (181 optical counterparts) and 
126 of them have unambiguous spectroscopic identification.
There are two spikes in the redshift distribution, predominantly populated 
by type-2 AGN but also type-1 AGN and X-ray normal galaxies: 
that at $z = 0.734$ is fairly narrow (in redshift space)
and comprises two clusters/groups of galaxies
centered on extended X-ray sources, the second one at $z = 0.674$ is 
broader and should trace a sheet-like structure. The type-1 and type-2
populations are clearly separated in X-ray/optical diagnostics involving
parameters sensitive to absorption/reddening: X-ray hardness ratio ($HR$), 
optical/near-IR colour, soft X-ray flux and optical brightness. 
Nevertheless, these two populations cover similar ranges of hard X-ray 
luminosity and absolute K magnitude, thus trace similar levels of
gravitational accretion. 
Consequently, we introduce a new classification based solely on X-ray 
properties, $HR$ and X-ray luminosity, consistent with the unified AGN 
model. This X-ray classification uncovers a large fraction of optically 
obscured, X-ray luminous AGNs missed by the classical optical classification.
We find a similar number of X-ray type-1 and type-2 QSOs 
($L_{\rm X}$(0.5-10 keV)$>10^{44}$ erg s$^{-1}$)
at $z > 2$ (13 sources with unambiguous spectroscopic 
identification); most X-ray type-1 QSOs are bright, R$ \lesssim 24$, whereas 
most X-ray 
type-2 QSOs have R$ \gtrsim 24$  which may explain the difference
with the CDFN results as few spectroscopic redshifts were obtained for
R$ > 24$ CDFN X-ray counterparts. There are X-ray type-1 QSOs down to 
$z\sim0.5$, but a strong decrease at $z<2$ in the fraction of luminous 
X-ray type-2 QSOs 
may indicate a cosmic evolution of the X-ray luminosity function of the type-2 
population. An X-ray spectral analysis is required to confirm this possible 
evolution. The red colour of most X-ray type-2 AGN could be due 
to dust associated with the X-ray absorbing material and/or a substantial 
contribution of the host galaxy light. The latter can also be important for
some redder X-ray type-1 AGN. There is a large population of EROs (R$-$K$>5$)
as X-ray counterparts and their fraction strongly increases with 
decreasing optical flux, up to 25\% for the R$ \geq 24$ sample.
They cover the whole range of X-ray hardness ratios,
comprise objects of various classes (in particular a high fraction of 
$z\gtrsim 1$ X-ray absorbed AGNs, but also elliptical and starburst 
galaxies) and more than half of them 
should be fairly bright X-ray sources ($L_{\rm X}$(0.5-10 keV)$>10^{42}$ 
erg s$^{-1}$). Photometric redshifts will be necessary 
to derive the properties and evolution of the X-ray selected EROs. 

\end{abstract}


\section{INTRODUCTION}

  Deep X-ray surveys indicate that the cosmic X-ray background
(XRB) is largely due to accretion onto supermassive black holes,
integrated over cosmic time. In the soft (0.5--2 keV) band more
than 90\% of the XRB flux has been resolved using 1.4 Msec
observations with {\em ROSAT} \citep{hasinger1998} and 1-2
Msec Chandra observations \citep{brandt2001a, rosati2002, brandt2002} 
and 100 ksec observations with XMM-Newton \citep{hasinger2001}. In
the harder (2-10 keV) band a similar fraction of the background
has been resolved with the above Chandra and XMM-Newton surveys,
reaching source densities of about 4000 deg$^{-2}$. Surveys in the
very hard (5-10 keV) band have been pioneered using BeppoSAX,
which resolved about 30\% of the XRB \citep{fiore1999}.
XMM-Newton and Chandra have now also resolved the majority
(60-70\%) of the very hard X-ray background.

Optical follow-up programs with 8-10m telescopes have been
completed for the {\em ROSAT} deep surveys and find
predominantly Active Galactic Nuclei (AGN) as counterparts of
the faint X-ray source population \citep{schmidt1998,
zamorani1999, lehmann2001}, mainly X-ray and optically unobscured
AGN (type-1 Seyferts and QSOs) and a smaller fraction of
obscured AGN (type-2 Seyferts). The X-ray observations have so
far been about consistent with population synthesis models based
on unified AGN schemes \citep{comastri1995, gilli2001}, which
explain the hard spectrum of the X-ray background by a mixture
of X-ray absorbed and unabsorbed AGN, folded with the corresponding
luminosity function and its cosmological evolution. According to
these models, most AGN spectra are heavily absorbed and about
80\% of the light produced by accretion will be absorbed by gas
and dust which may reside in nuclear starburst regions that
feed the AGN \citep{fabian1998}. However, these models are far from
unique and contain a number of often overlooked assumptions, so their
predictive power remains limited until complete samples of
spectroscopically classified hard X-ray sources are available.
In particular, they require a substantial contribution of
high-luminosity absorbed X-ray sources (type-2 QSOs), which so
far have only scarcely been detected. The cosmic history of
obscuration and its potential dependence on intrinsic source
luminosity remain completely unknown. \citet{gilli2001} e.g.
assumed a strong evolution of the absorbed/obscured fraction (ratio of
type-2/type-1 AGN) from 4:1 in the local universe to much larger
fractions (10:1) at high redshifts \citep[see
also][]{fabian1998}.  The gas-to-dust ratio in high-redshift,
high-luminosity AGN could be completely different from the
usually assumed Galactic value due to sputtering of the dust
particles in the strong radiation field \citep{granato1997}.
There could thus be objects which are heavily absorbed at X-rays
and unobscured at optical wavelengths.

After having understood the basic contributions to the X-ray
background, the general interest is now focussing on
understanding the physical nature of these sources, the
cosmological evolution of their properties, and their role in
models of galaxy evolution. We know that basically every galaxy
with a spheroidal component in the local universe has a
supermassive black hole in its centre \citep{gebhardt2000}.  The
luminosity function of X-ray selected AGN shows strong
cosmological density evolution at redshifts up to 2, which goes
hand in hand with the cosmic star formation history
\citep{miyaji2000}.  At the redshift peak of optically selected
QSOs, around $z$=2.5, the AGN space density is several hundred times
higher than locally, which is in line with the assumption that
most  galaxies have been active in the past and that the feeding
of their black holes is reflected in the X-ray background. While
the comoving space density of optically and radio-selected QSOs
has been shown to decline significantly beyond a redshift of 2.5
\citep{schmidt1997, fan2001, shaver1996}, the statistical quality
of X-ray selected high-redshift AGN samples still needs to be
improved \citep{miyaji2000}.  The new Chandra and XMM-Newton
surveys are now providing strong additional constraints.

Optical identifications of the deepest Chandra and XMM-Newton
fields are still in progress, however, a mixture of obscured and
unobscured AGN with an increasing fraction of obscuration at
lower flux levels seems to be the dominant population in these
samples \citep{fiore2000, barger2001a, tozzi2001,
rosati2002, stern2002}. First examples of the long-sought class
of high-redshift, radio-quiet, high-luminosity, heavily obscured active
galactic nuclei (type-2 QSO) have also been detected in deep
Chandra fields \citep{norman2002, stern2002} and in the
XMM-Newton deep survey in the Lockman Hole field
\citep{hasinger2002}.

In this paper we report on our optical identification
work in the Chandra Deep Field South, which, thanks to the
efficiency of the VLT, has progressed to the faintest magnitudes
among the deepest X-ray surveys.

\section{THE CHANDRA DEEP FIELD SOUTH (CDFS)}

The Chandra X-ray Observatory has performed deep X-ray surveys
in a number of fields with ever increasing exposure times
\citep{mushotzky2000, hornschemeier2000, giacconi2001, tozzi2001,
brandt2001a} and has completed a 1 Msec exposure in the Chandra
Deep Field South, CDFS \citep{giacconi2002, rosati2002} and a 2
Msec exposure in the Hubble Deep Field North, HDF-N
\citep{brandt2002}.  The Megasecond dataset of the CDFS is the
result of the coaddition of 11 individual Chandra ACIS-I
exposures with aimpoints only a few arcsec from each other. The
nominal aim point of the CDFS is $\alpha=3:32:28.0$,
$\delta=-27:48:30$ (J2000).  This field was selected in a patch
of the southern sky characterized by a low galactic neutral
hydrogen column density, $N_{\rm H}=8\times 10^{19} \cm^{-2}$, and a lack
of bright stars \citep{rosati2002}.

\section{OPTICAL IDENTIFICATIONS IN THE CDFS\label{optid}}

Our primary optical imaging was obtained using the FORS1 camera
on the ANTU (UT-1 at VLT) telescope. The
R band mosaics cover $\sim$360 arcmin$^2$ to depths
between 26 and 26.7 (Vega magnitudes). This data does not cover
the full CDFS area and must be supplemented with other
observations (see Figure \ref{position}). The ESO Imaging Survey (EIS) has covered this
field to moderate depths (5 $\sigma$ limiting AB magnitudes of 26.0, 25.7,
26.4, 25.4, 25.5 and  24.7 in U$^\prime$, U, B, V, R and I, respectively)
in several bands \citep{arnouts2001,
vandame2001}. The EIS data has been obtained using the Wide
Field Imager (WFI) on the ESO-MPG 2.2 meter telescope at La
Silla.  The positioning of the X-ray sources is better than 0.5\arcsec
\citep{giacconi2002} and we readily
identify likely optical counterparts in 85\% of the cases.

\begin{figure}
\plotone{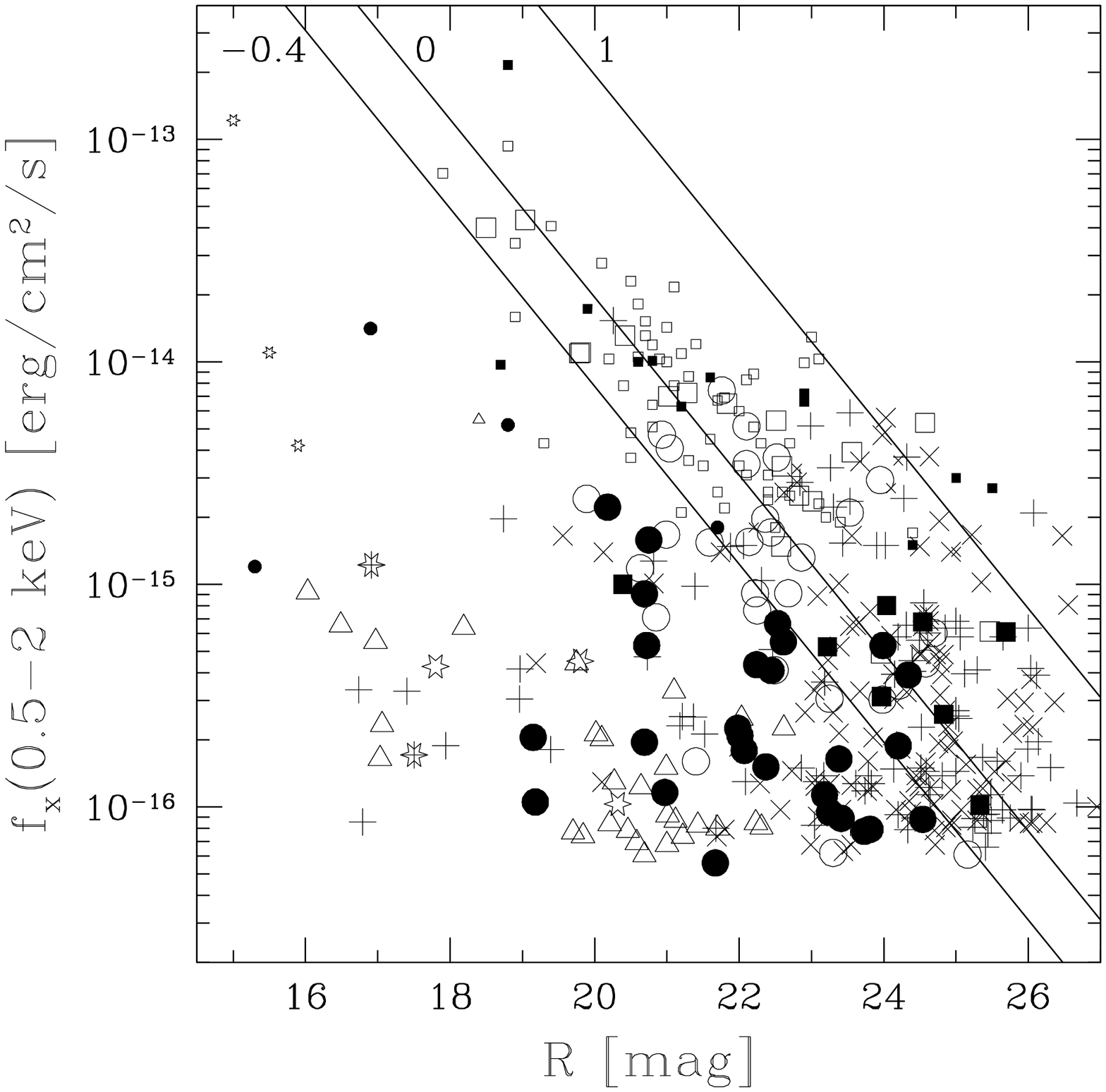}

\caption{X-ray flux in the 0.5-2 keV band versus R-band magnitude for the
CDFS-sources (larger symbols) and the ultradeep ROSAT survey in
the Lockman Hole (smaller symbols). Objects are marked according to
their X-ray classification for the CDFS sources 
and the original classification for the Lockman Hole sources:
squares correspond to QSOs, circles to AGN and triangles to galaxies.
Type-1 and type-2 AGNs/QSOs have empty and solid symbols, respectively.
Stars are marked with star symbols.  `X' symbols refer to 
spectroscopically not securely identified CDFS counterpart candidates which
we observed in our program, `+' symbols mark objects we did not observe.
The solid lines correspond to an X-ray to optical flux ratio index
(as discussed in Section \ref{optid}) of 1, 0 and $-$0.4.\label{fxR}}

\end{figure}

\begin{figure}
\plotone{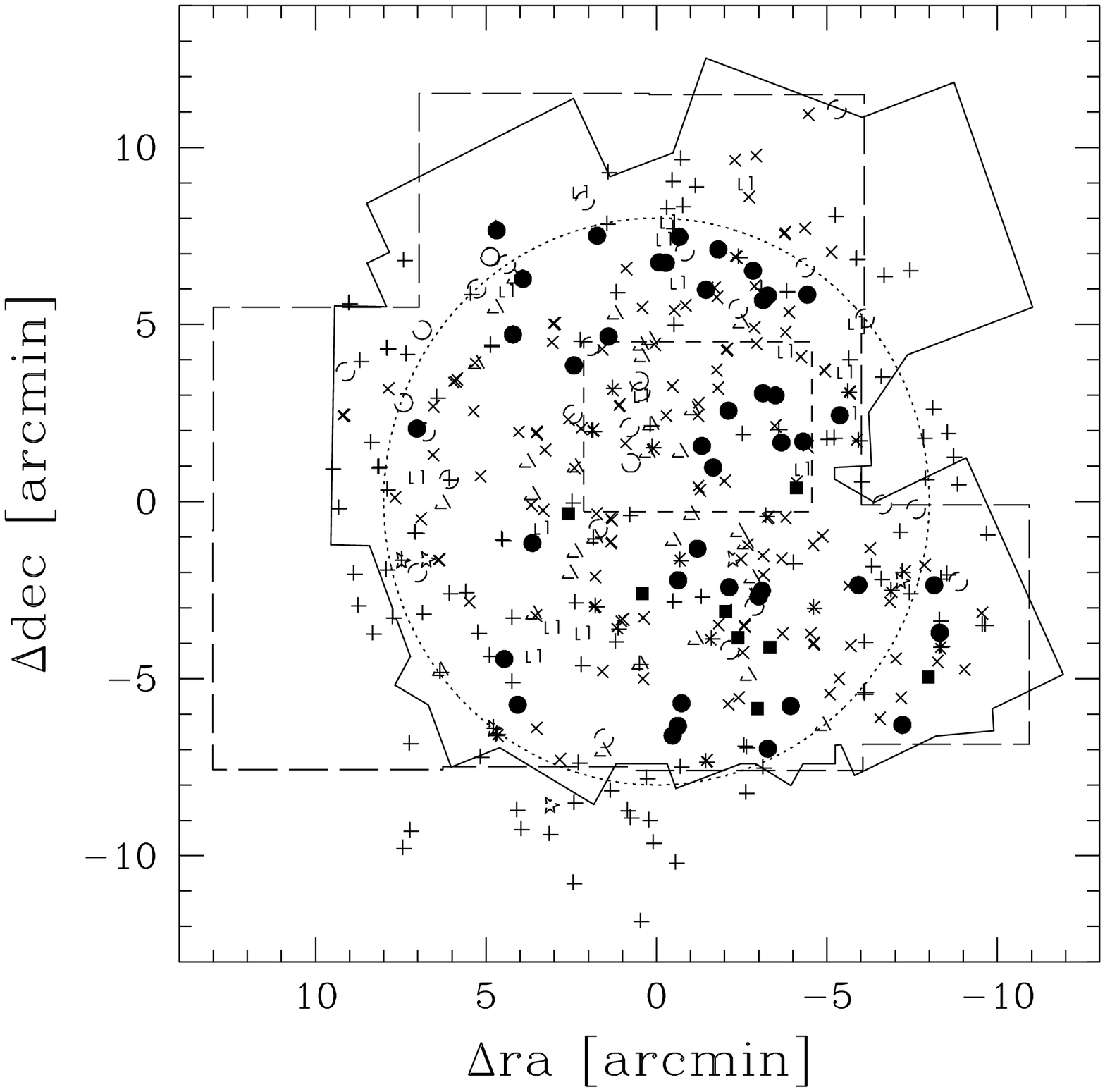}

\caption{Spatial distribution of the X-ray sources with and without
pectroscopic observations. 
Object classes are marked as in Figure \ref{fxR}. The area covered
by our spectroscopic masks (solid line), the imaging survey (long dashed
line), the K20 survey area (short dashed line) and an 8 arcmin radius
circle, centered on the nominal CDFS
pointing (3:32:28.0 -27:48:30 J2000; dotted line) is also shown.
\label{position}}

\end{figure}

Figure \ref{fxR} shows the classical correlation between the 
R-band magnitude and the soft X-ray flux of the CDFS sources.
The objects are marked according to their classification 
(see below). By comparison with the deepest ROSAT survey in the 
Lockman Hole \citep{lehmann2001},  the Chandra data extend the
previous ROSAT range by a factor of  about 40 in X-ray flux and to
substantially fainter optical magnitudes. While the bulk of the
type-1 AGN population still follows the general correlation
along a constant $f_X/f_{opt}$ line, the type-2 AGNs cluster at
higher X-ray-to-optical flux ratios. There is also a population
of normal galaxies emerging at low fluxes (thus discovered in
the Chandra and XMM era).

To be consistent with the already published deep ROSAT catalogs
\citep{lehmann2001},
we used a modified version of the X-ray to optical flux ratios:
\begin{equation}
log_{10}(f_x/f_o)\equiv log_{10}(f_{0.5-2\keV}/f_R)\equiv
log_{10}(f_{0.5-2\keV})+0.4 R + 5.71,
\end{equation}
where the flux is measured in erg cm$^{-2}$ s$^{-1}$ units in the 0.5-2
keV band and R is in Vega magnitudes. The slight change in the
normalization \citep{maccacaro1988} is motivated by the
significantly narrower X-ray energy band used (the original
definition was based on the {\em Einstein} medium sensitivity
survey band, 0.3-3.5 keV), which introduces a factor of 1.77
decrease in the flux for objects with a spectral energy index of $-$1
(classical type-1 AGN) and the use of the R-band instead of V
(here we assumed a V$-$R color of 0.22, typical
value for galaxies).

To use this new X-ray to optical flux ratio definition for source
classification, we also had to convert the canonical
ranges \citep{stocke1991} to our new system. The new ranges for
different classes of objects are shown in
Table \ref{fxfo_classes}. To calculate the new ranges of the
X-ray to optical flux ratios, we assumed typical X-ray spectra
for each class and calculated the shift in the X-ray flux due to
the narrower energy band: a power law with a photon index of
$\Gamma$=1-2.7 for AGN, a power law with a photon index of
$\Gamma$=1-2 for BL Lac objects, a Raymond-Smith model with
$kT=$2-7 keV, abundances of 0.1-0.6 and redshifts of
$z=$0-1. For stars and supernova remnants we used
Raymond-Smith models with energies of $kT$=0.5-2 keV, for
X-ray binaries powerlaw models with photon index
$\Gamma$=1-2.  For galaxies, we adopted a somewhat ad
hoc shift of 0.1-0.3 in the logarithm of the flux due to the
different energy bands.  This choice was motivated by examining
different models for galaxies (warm and hot plasma mixture,
powerlaw like emission from X-ray binaries, typical supernova
remnant spectra, etc.).


For the shift in the optical flux (using R-band instead of the
canonical V-band) we assumed typical values for each class.

The resulting ranges of the X-ray-to-optical flux ratios are
shown in Table \ref{fxfo_classes}.  As can be seen from the
table, the new X-ray-to-optical flux ratio is not significantly
different from the canonical one. The typical ranges are a bit
wider, but this just a consequence of {\em converting} the
ranges instead of directly determining it from large surveys.
With our new normalization, we can use the original ranges
\citep{stocke1991} to make an educated guess on the
galactic/extragalactic nature of objects.

\section{TARGET SELECTION}

Target selection was primarily based on our deep VLT/FORS imaging 
data \citep{giacconi2002}, reaching a depth of R $\sim26.5$. In regions 
not covered by this VLT/FORS deep imaging, we used somewhat shallower 
VLT/FORS imaging in the R-band obtained as part of the survey.

Possible optical counterparts of X-ray sources were selected
based on the estimated astrometry error of the X-ray object (for
a relatively bright point source at zero off-axis angle the
astrometry rms error is $\sim 0\farcs5$). We used
the automatically generated optical
catalog, however, {\em every} object was visually inspected for
deblending problems and artefacts.

The surface density of our X-ray objects is very well suited to
MOS spectroscopy with FORS/VLT. We could fill a
large part of the masks with program objects and it was quite
rare that we had to choose between multiple optical counterpart
candidates within
the geometrical constraints of the instrument.  As a
consequence, our target selection is nearly unbiased. The only
selection effect that should be considered was related to
objects with multiple counterpart candidates. In these
cases we usually selected the object in the appropriate
magnitude range for the particular mask, but in general we tried
to revisit these objects -- unless the first one turned out
to be clearly the counterpart.

We also took advantage of the extremely high accuracy of the
robotic masks: in some cases, we reconfigured {\em some} of the
slits between read-outs, without changing the telescope pointing
to observe many (brighter) optical counterparts. This way, 
the integration time
on bright objects could be shortened and we could use the remaining
time on a different object, while maintaining longer integration
times for the faint ones.

During our last two runs (in November and December 2001) we were also using
the prefabricated masks (MXU mode -- only available for FORS2), as opposed to
movable robotic slitlets (MOS mode). For our survey, the only important 
difference between the two modes is more freedom in the placement of
slits in MXU mode. This improved our observing efficiency in the
later phase of the survey, where we concentrated on fainter objects
(with a higher surface density).

\subsection{\it The Reliability of the Target Selection}

\begin{figure}
\plotone{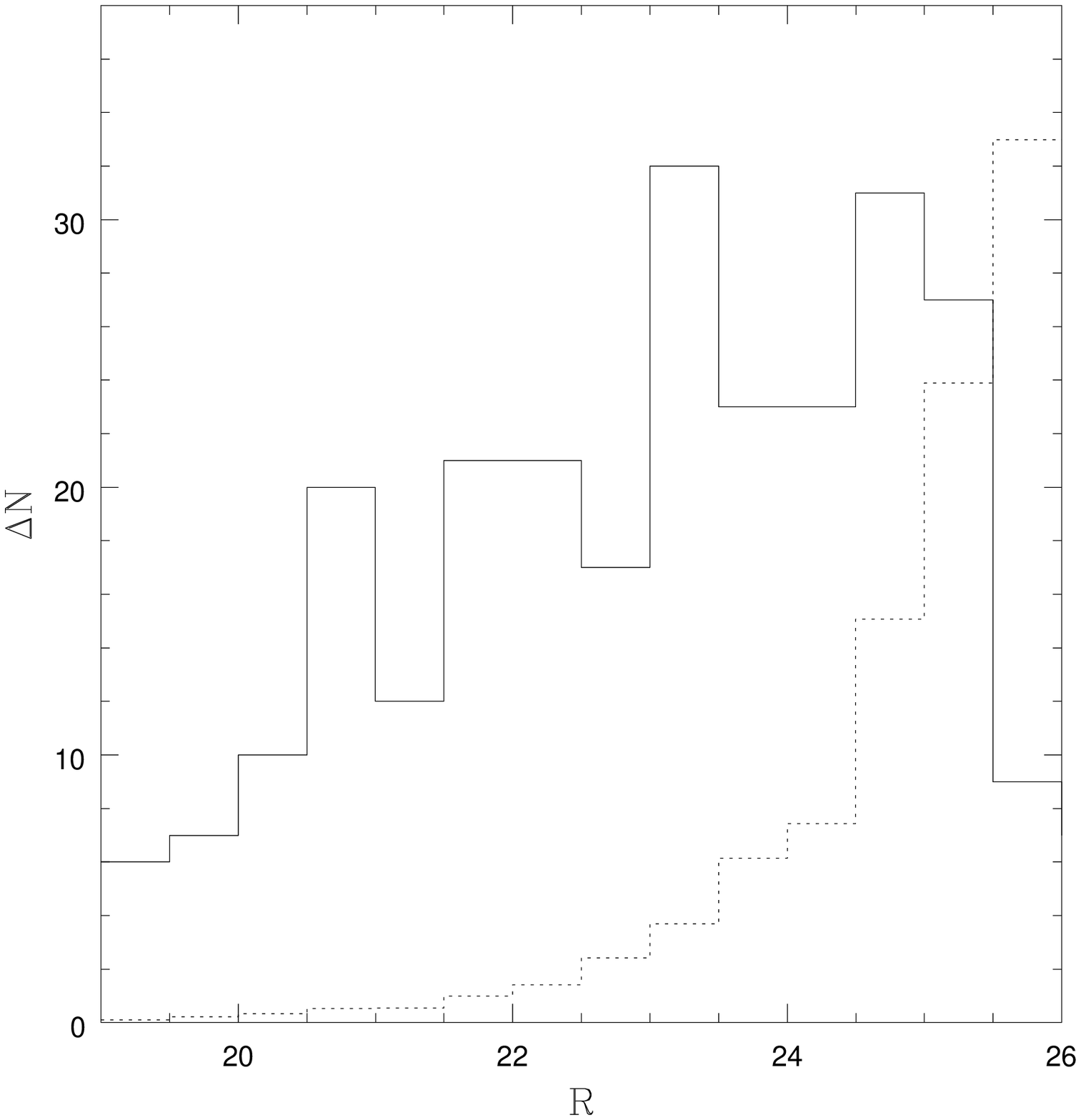}
\caption{The R-band magnitude distribution of the selected (primary) optical
counterparts of the X-ray sources in our survey (solid line). For comparision, 
we also show (dotted line) the expected distribution of random field galaxies
normalized to the total area of the error circles of our X-ray sources,
based on galaxy number count measurements \citep{metcalfe2001,jones1991}.
\label{magdist}}
\end{figure}

The reliability of X-ray follow-up surveys using optical (or near infrared)
spectroscopy hinges on matching the X-ray source to the {\em right} optical
object. This is primarily done through astrometry. Just how reliable are
these identifications? Using deep galaxy number counts \citep{metcalfe2001},
we expect roughly 0.02 galaxies in every square arcsec area 
that are brighter than R $\sim26$. Considering our best $3\sigma$
astrometry error ($1\farcs5$), we expect $\sim0.15$ {\em field} galaxies
to fall within our error circle -- even in the best, zero off axis angle
case. In other words we expect one {\em false} candidate for every 
seventh X-ray object at R $<26$. Considering the roughly 250 X-ray
sources we observed, we expect that for at least 35 of them, there
{\em will be} a completely unrelated faint galaxy, even in an error circle
of $1\farcs5$. Fortunately, the X-ray counterpart candidates typically 
have much brighter magnitudes (see Figure \ref{magdist}). At these
brighter magnitudes the probability of field galaxy contamination is
much lower, so we should only worry about contamination for very
faint (R=25-27) objects.

As our astrometric accuracy heavily depends on the
signal-to-noise ratio of the object (i.e. objects with low
photon counts are centered with lower accuracy) and the off-axis
angle of the object (there is a significant degradation of the
PSF of increasing off-axis angles), the total area covered by
the sum of the error circles is quite large, around 3900
arcsec$^2$. In Figure \ref{magdist}, we show the magnitude
distribution of our selected primary optical counterparts and
the expected magnitude distribution of random field galaxies
over this area, based on galaxy number counts
\citep{metcalfe2001,jones1991}.  Contamination by random field
galaxies becomes a serious problem beyond R $\sim24$ and they
start to dominate beyond R $\sim26$, the practical limit of our
imaging survey.

Therefore, extra caution is required in making sure that the
{\em right} optical object is identified as the counterpart.
This is not always trivial as the optical spectra do not always
show clear signatures of active nuclei (AGNs).  In some cases we
had to observe every object in the error circle. Fortunately
this turns out to be feasible. At R $\sim24$ and fainter,
deblending is not a serious challenge (using both automated and
visual tests). At brighter magnitudes, where deblending would be
near impossible (e.g.  detecting a R $\sim25$ X-ray object in the
halo of a R $\sim19$ galaxy), the probability of field galaxy
contamination is negligible.  Stellar contamination is
negligible at our high galactic latitude.

It is also important to point out that these estimates of field galaxy 
contamination are for the probability of finding an unrelated object
in our X-ray error circle. We can also ask a technical question: what is the
probability of finding a field object on a slit? Taking a
$20\arcsec\times2\arcsec$ area (the typical slit length in FORS-1 is 
around 20$\arcsec$), we expect to
find a R $<23$ galaxy in 5\% of the slits and we expect statistically
a field galaxy with magnitude R $<25$ in every second slit. This means that
one has to be extremely careful in the data reduction and do a very careful
book keeping in the process.

\section{OBSERVATIONS AND DATA REDUCTION}

Data were obtained during 11 nights in 2000 and 2001. A summary
of the observations is presented in Table \ref{tblobs}. All
observations were using the `150I' grism (150I+17 in  FORS-1 and
150I+27 in FORS-2). These grisms provide a pixel scale
(dispersed) of 280\AA/mm, or roughly 5.5\AA/pixel.  The nominal
resolution of the configuration is
$R=\lambda/\Delta\lambda$=230, which corresponds to roughly
20{\AA} at 5600{\AA}.  The pixel scale of these instruments is
0.2$\arcsec$/pixel, so there is no significant degradation of the
resolution due to the finite slit width.

In the initial phase of our survey, we exclusively used low
resolution multiobject spectroscopy with varying integration
time. This strategy maximizes the number of observed objects and
provides a (nearly) full spectral coverage for every exposure.
This is clearly a trade off, as we then get a significantly
lower S/N spectrum for the individual objects, compared to
higher resolution long-slit spectroscopy based on photometric
redshifts, but the latter technique was deemed to be
prohibitevely expensive in observing time in the initial phase
of our project.

As our goal was to observe as many objects as possible, we used
non standard order separation filters (either no filter,
or the GG-375 filter, which cuts out light bluer
than $\sim$3750\AA). It was thus possible to cover a very wide
spectral range in a single exposure (in the standard configuration
the order separation filter that cuts the light blueward of
5900\AA, thus the whole spectral range can only be covered in
two exposures).

\subsection{\it Data Reduction}

Data were reduced by our own semi-automatic pipeline built on top of
IRAF. In general we followed standard procedures, but
had to deviate slightly in several cases to accomodate particularities 
of the FORS instrument and do a very rigorous book-keeping. In the 
following sections, we enumerate these changes.

\subsection{\it Bias, Overscan and Trim Correction}

The FORS CCD's have in principle 4 read-out modes: high and low gain
and one and four amplifier modes. To avoid serious complications, we
only used the high gain/one amplifier read-out mode for our spectroscopic
observations. This decision resulted in a slightly larger overhead, but
this was deemed negligible considering our long integration times,
compared to the challenges posed by reducing a 4 amplifier read-out mode
spectroscopic observations, where we would have to calibrate the gain
of each amplifier very accurately (so we do not introduce artifical
features in the spectra).

A sufficient number of full frame bias exposures were taken during each
run (typically around 20 per run). These were individually overscan
corrected and trimmed. The resulting (bias) frames were averaged with suspect
pixels (too high or too low values) filtered out to generate the master
bias frame. In each case we verified that the bias frame does not change
significantly from night to night within a run.

A slight complication was posed by our spectrophotometric
standard observations. These frames were also using one
amplifier/high gain, but (to save some time) only 500 rows were
read out (centered on the standard star). Since ESO does not
provide an under/overscan region for windowed frames, we took a
sufficient number (typically 10) of bias frames in this
configuration. Naturally (lacking under/overscan region) these
frames were {\em not} overscan corrected, nor trimmed.  Instead,
they were averaged to create a master bias frame, which {\em did}
include the artificially introduced bias level. We checked the
individual frames and confirmed that the variation of this
artifical bias level is negligible for these very high
S/N frames.

After creating the full and windowed bias frames, all
object and calibration (flat and arc) frames were overscan corrected
and trimmed (except the windowed frames) and zero subtracted.

At this point we applied a shift in the dispersion direction,
based on the slit position, to bring (very crudely; within 10
pixels or 50\AA) the observations on a similar wavelength scale.
We also inserted gaps in the spatial direction between the
neighbouring slits to reduce the risk of contamination between
slits.  These two steps are purely practical, but make
bookkeeping significantly easier.

\subsection{\it Flatfielding}

In this processing step, we had to tackle three main issues:

The first one is an inherent complication in the FORS
instruments. Due to the mechanical construction of the robotic
slit masks and the location of the flat-field lamps, flat-field
exposures show higher flux levels in a few rows at the upper or
lower edge of the slit. To correct for this effect, there are
two sets of flat-field lamps in the instrument. We took a
sufficient number of flat-field exposures using both sets of
lamps. We generated merged flats independently for each lamp set
and generated the final flat-field frame by taking the smaller
pixel value in the two frames. As the reflections from the two
lamp sets do not overlap, this feature can be fully removed.

The second issue is a consequence of our unusual observing
strategy.  In some cases (due to geometric constraints imposed
by the robotic slit masks) we could not target very faint
objects with a particular slit, but we had several bright
candidates available.  In these cases, to maximize efficiency,
we reconfigured these slits between readouts so that all bright
candidates were observed, while faint objects targeted with
other slits were observed with a longer integration time. Due to
the extremely high mechanical stability of the FORS instruments,
this strategy is very safe.  As the sensitivity variation
between pixels is potentially color dependent, we decided to
generate flat-field frames for each mask.  This may not be the
optimal strategy since for the slits that are in the same
position in two masks, we could use more exposures, thus to
create a more accurate flat-field. This alternative strategy
would be too complex and the resulting data quality improvement is very
marginal, consequently we decided {\em against} it.

The last major issue is due to the extremely wide spectral coverage
used. As our intention was to correct {\em only} for the pixel-to-pixel
sensitivity variations, we had to generate a normalization
image (a combination of the flat-field lamp spectrum and the overall
quantum efficiency of the system as a function of wavelength and spatial
position). For high resolution (and smaller wavelength coverage)
observations, this is often achieved by collapsing in the spatial
direction and fitting a function in the dispersion direction. Unfortunately,
this technique proved to be impractical for us. The main problem was that
we were unable to find an ansatz function that could reproduce the very
sharp cutoffs at both ends (due to either the order separation filter or
the natural cut-off of the CCD detector) without introducing artifical
structure on intermediate scale. An additional complication was that
the internal flat-field lamps did not illuminate the slits homogenously
-- there is a slight gradient in the spatial direction. Therefore,
after a slight smoothing of the flat-field exposures, we created the
normalization image by a linear or (for very long slits) a second
order polynomial fit in the spatial direction. Each flat-field exposure
was divided by this normalization frame, thus creating a `true' flat-field
frame, which only contains pixel-to-pixel sensitivity variations. In regions,
where the signal was too low, the flat-field was artificially set to one 
(to avoid the introduction of too high photon noise).

After these steps, the individual, normalized flat-field frames were
merged, eliminating the effect of light reflection on the slit edges.
Both science and wavelength calibration frames for a given
mask were divided by the resulting master flat-field frame.

\subsection{\it Sky Subtraction}

The sky background was estimated in each column by a linear fit
(for longer slits) or just calculating the average (shorter
slits) in each column of each slit, rejecting too high pixels
(i.e. the targeted object) and subtracting the result. It is
important to note that we did {\em not} correct for the very
slight curvature of the dispersed spectra on the CCD in this
step. With the FORS instruments, this strategy works quite well
(as opposed to LRIS on the Keck telescope). Significant sky
residuals are only present around the very bright, narrow sky
lines -- where sky subtration is doomed anyway due to pixel
saturation.

This procedure works only for our typical faint objects.
Extremely bright objects can illuminate the
whole slit, thus making correct sky subtraction impossible. Fortunately,
in those (very few) cases identification was still possible due to the
extremely high object signal.

\subsection{\it Fringe Removal}

In some cases (especially in MXU masks), we could take advantage of our
dithering strategy to reduce further the effect of fringing and
the sky residuals. As neither the fringe pattern nor the sky residuals
are significantly affected by the small (spatial) offsets of the telescope,
we could, in some cases (with sufficient number of exposures in a given
mask) exclude (most of) the object signal and create a fringe/sky residual
template for each slit. Subtracting this from the frames resulted 
in an improved signal-to-noise ratio for the object spectra. Depending
on the seeing conditions and the dithering offsets used, not all object
signal was perfectly removed, thus the extracted spectra significantly
underestimated the real spectra. As our primary goal was object identification,
not spectrophotometry, this was an acceptable trade-off.

\subsection{\it Coadding the Frames}

After sky subtraction, all the slits were visually inspected to
verify that the object is indeed in the 'good' region of the
slit. This step was necessary since the applied small spatial
offsets between the science exposures can result in objects
falling too close to the slit edge (MOS blade corners are round,
thus the slit is not usable there) or falling completely outside
the slit.

After this visual screening, the {\em spatial} offset between
different exposures of the same object was caculated based on
the {\em world coordinate system} (WCS) information stored in
the frame headers. The individual exposures were coadded
(including the rejection of suspicious pixels or cosmic ray
hits) after applying these spatial shifts. We only shifted the
frames in the spatial direction and only by integer number of
pixels. As the objects were sufficiently well sampled (the pixel
scale was significantly smaller than the seeing), this step
resulted in nearly negligible bluring of the spectra, while
preserving the statistical properties of the exposures.

\subsection{\it Extraction}

Even though for {\em sky subtraction} we could safely ignore the
slight curvature of dispersed spectra on the CCD, for the
extraction of the object signal this is no longer possible.
Therefore, we estimated the object position on the detector by
collapsing at least 30 columns (more for really faint objects)
in the dispersion direction and measuring the object center in
the resulting profile. The object position was fitted typically
with a second order polynomial as a function of column
(wavelength).

Then an aperture {\em width} was visually determined. Except in
special cases (e.g. blended objects), our aim was to include
most of the object signal without adding too much sky (to
maximize the signal-to-noise ratio).  A one dimensional spectrum
was obtained using the `optimal extraction' method of IRAF.
This procedure calculates a weighted average in each column,
based on both the estimated object profile and photon
statistics.

\subsection{\it Wavelength Calibration}

Wavelength calibration was based on (daytime) arc calibration frames, using
four arc lamps (a He, a HgCd and two Ar lamps) which provide a
sufficient number of lines over the whole spectral range used (3889--9924\AA).

The exact same aperture that was used for the science object was
used for the arc frames. The resulting lines were first
identified automatically.  These identifications were then
visually verified, and quite often significantly improved. In
most cases, around 20 lines were located in the 3889--9924{\AA}
range, and fitted by a forth order polinomial,
with a typical rms accuracy of 1{\AA} or better.  This
accuracy is close to that expected from the
nominal resolution of the instrument in our configuration.
The object spectra were then wavelength calibrated and
subsequently rebinned to obtain spectra with a linear
wavelength scale.

We also examined the stability of the instrument by repeating the daytime
calibrations on different days. No noticeable change was detected. In 
addition, we verified the wavelength calibration by checking the position
of narrow skylines in science exposures -- no discrepancy was found within our error estimates.


The wavelength
calibration may not be accurate in the range outside the two
extreme arc lines identified.  As the FORS instruments use a
grism, we had to resort to high order polynomial fits,
which become unreliable when extrapolating the
wavelength solution. In most cases, this is not an
important issue, but there were a few unfortunate cases where
major object features fell into unreliable regions
(typically if the spectra were cut short on the blue side due to
the position of the slit).

\section{FLUX CALIBRATION}

In this step, we {\em nearly} followed standard practices.
The only significant necessary change arrised due to our choice of
a non standard instrument configuration, namely not using the {\em
right} order separation filters. Consequently, we nearly doubled
our efficiency (taking only one exposure per object), but we
then had to correct for second order diffracted light.

It is important to point out that we have to correct for this effect
both for the science and the spectrophotometric standard
observations. 

\subsection{\it Second Order Diffraction\label{2ndtheory}}

The first step was to determine the nature of the second order
contamination.  As this contamination affects the red part of
the spectrum, where there are typically many lines present
already from first order diffraction (in both arc and sky
exposures), we used a special set of calibration frames:
a 1.3 arcsec wide long slit, the standard arc lamps
(He, HgCd, Ar) and a set of (dispersed) exposures through all available
broad-band filters (U, B, V, R and I) as well as without any filter.
Using the exposure without filter, we established the {\em first order}
wavelength solution of this configuration. We verified that the use of
the Bessel filters does {\em not} introduce any noticable shift
in this solution. We were then able to
identify the second order lines in the exposures taken through the
broad-band filters. These identifications are shown in Table \ref{2ndorder}.

The comparison between the apparent fluxes in first and second order diffracted
lines indicates that second order diffraction {\em can} be very strong,
as much as 30\% of the first order strength (especially in the blue
part of the spectrum). This effect is made seriously worse by the
quantum efficiency of the CCD.
The overall quantum efficiency of the system peaks around 6000{\AA} and
declines relatively rapidly (see also Section \ref{specphotstd}).
Because of this, a relatively weak second
order contamination may become the dominant signal beyond 9000\AA{} -- 
due to to the much higher sensitivity of the pixels to these photons.
For blue objetcs (for example spectral photometric standards), this
problem is even worse: second order contamination can already start
at 7000\AA (due to the high UV flux of the object) and can contribute
over 30\% to the observed flux.

To correct for the second order diffraction, we first had
to model it. Based on the identified arc lines, we adopted
a second order wavelength solution in a linear form:
\begin{equation}
\lambda=2.106\Lambda-723\hbox{\AA},
\end{equation}
where $\Lambda$ is the real wavelength of the feature and $\lambda$ is its
apparent wavelength position observed in second order.

It should be noted that, as opposed to {\em grating} spectrographs
where the coefficient is practically two and the shift is very small,
a few times 10\AA{} \citep{gutierrez1994}, FORS, which uses {\em grisms},
is significantly different, which makes the detection of second order
diffraction harder, unless one takes the appropriate calibration
data sets.

We assumed that the measured signal, $d(\lambda)$ is
\begin{equation}
d(\lambda)=f(\lambda)s(\lambda)+c(\Lambda)f(\Lambda),
\end{equation}
where $s(\lambda)$ is the overall quantum efficiency of the system,
$\Lambda$ is the
real, physical wavelength of features detected at $\lambda$ in second order,
$c(\Lambda)$ is the strength of the second order folded into the sensitivity
function at $\Lambda$ and $f(\lambda)$ is the real spectrum of the object.

Since we can not derive {\em two} functions
($s(\lambda)$ and $c(\Lambda)$) from a single measurement,
either we determine the sensitivity, $s(\lambda)$, independently
(e.g. observing a standard star through different order
separation filters) or we use two independent measurements
from observing two different standard stars. As the first option implies
the introduction of an additional optical element (the filter),
which can affect the strength of the second order diffracted
signal (in fact comparing the flux ratios of the 3650.1\AA{} arc
line in Table \ref{2ndorder} is a strong indication for this to
be the case), we selected the latter aproach. We observed two
standards with very different spectral shapes, LTT-3218
\citep[ a relatively red DA6 white dwarf]{hamuy1992,hamuy1994} and
HD49798 \citep[ a blue sdO6 subdwarf]{turnshek1990}.

Given two different standards ($f_1(\lambda)$ and $f_2(\lambda)$), but 
identical instrument setups ($c(\Lambda)$ and $s(\lambda)$), we can write

\begin{equation}
c(\Lambda)={d_1(\lambda)-f_1(\lambda)s(\lambda)\over f_1(\Lambda)}=
{d_2(\lambda)-f_2(\lambda)s(\lambda)\over f_2(\Lambda)},
\end{equation}
which we can solve for $c(\Lambda)$:
\begin{equation}
c(\Lambda)={d_1(\lambda)\over f_1(\Lambda)}\left(1-
{\alpha(\Lambda)-\beta(\lambda)\over\alpha(\Lambda)-\alpha(\lambda)}\right),
\end{equation}
where $\alpha(\lambda)=f_2(\lambda)/f_1(\lambda)$ (known a'priori) and
$\beta(\lambda)=d_2(\lambda)/d_1(\lambda)$ (known from observations).

The derived $c(\Lambda)$ indicates that the contamination is
completely negligible up to 6300\AA. Up to 7500\AA{} it is
somewhat stronger, but typically still negligible as this
wavelength corresponds to up to 3800\AA{} in first order, where
the CCD is very inefficient.
Between 7500 and
9000\AA{}, the effect is strong (12\% to 3\% of the first order
instrumental flux shows up in second order). Depending on the
object type (whether it is a blue or red object) and the
wavelength of interest (as this result should be folded with the
system quantum efficiency, which is increasing with wavelength
for second order diffracted photons and decreasing for first
order diffracted photons in this range) this may or may not be a
strong effect -- this decision should be made for each observing
program.  Beyond that this effect could not be estimated
as one of our standards is measured only to 8700\AA.
As in this range the CCD QE is dropping very sharply, while the
first order QE is very high, second order contamination {\em
must} become the dominant source of signal at some wavelength.

\begin{figure}
\plotone{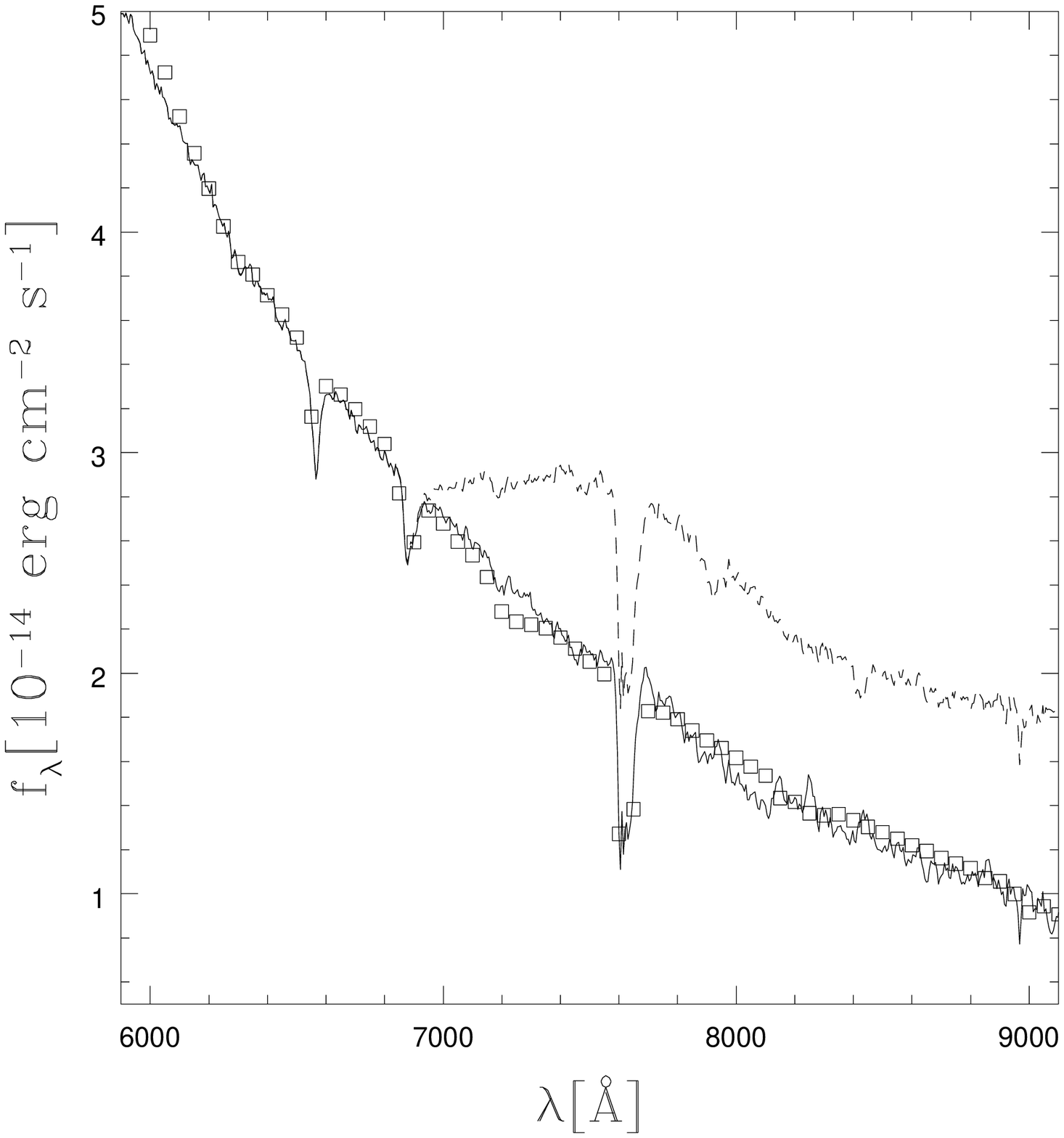}
\caption{Flux calibrated spectra of a spectrophotometric standard star,
Feige110 \citep{hamuy1992,hamuy1994,oke1990}. The dashed line is the
measured flux without removing the second order contamination, solid line
is the spectra {\em after} the correction. Empty squares are the 
real flux values from literature.\label{secondorderplot}}.
\end{figure}

The effect of the second order diffraction is demonstrated in Figure
\ref{secondorderplot}, using a blue spectrophotometric standard star,
Feige110 \citep{hamuy1992,hamuy1994,oke1990}.

\subsection{\it Overall Quantum Efficiency of the System\label{specphotstd}}

The overall throughput of the system was determined using a set of
spectro photometric standard stars:  LTT-377, LTT-3218,
LTT-7379, LTT-7987, LTT-9239 \citep{hamuy1992,hamuy1994}, HD49798
\citep{turnshek1990} and Feige110
\citep{hamuy1992,hamuy1994,oke1990}.  These objects were observed
repeatedly over a wide range of airmasses during each run, using
a simulated very wide ($\sim 5\arcsec$) long slit, using the
robotic mask facility of the FORS instruments (combining 3
slits).

The observations were reduced nearly the same way as
science observations. Second order diffraction was removed as
outlined in Section \ref{2ndtheory}. The measured spectra (in instrumental
units) were compared to the published physical spectra -- excluding
regions with sharp features in the objects and sharp telluric absorption
features in the atmosphere. A smooth sensitivity curve was
fitted to the data points. As we were using about the whole
wavelength range of the CCD, this fit was done in four parts:
3000-4000{\AA} (or 3500-4000{\AA} if the OG375 order separation 
filter was used during the run), where the throughput raises very
sharply and only 
a 10\% accuracy was achieved, between 4000-5000{\AA}, where the
throughput is still rising fast (about 4\% accuracy),
between 5000-8000{\AA} (2\% accuracy)
and finally between 8000-9500{\AA}, where the accuracy drops again
to around 10\%. These four data sets were used to construct the
sensitivity curve for each observing run and each configuration.

As the Paranal Observatory does not have sufficient data collected to
measure an accurate spectroscopic extinction curve, we used the
curve published by the Cerro Tololo Inter-American Observatory (CTIO),
after verifying, using our standard observations, that the curve
is very close to that estimated for Paranal.

This calibration was applied to all measured program spectra. We
also corrected for atmospheric extinction using the CTIO
extinction curve.  The accuracy of the resulting flux calibrated
(but not absolute calibrated -- see below) spectra is
mainly constrained by the signal-to-noise ratio of the
object signal in the 4000-8500{\AA} range (the inaccuracy due to the
sensitivity function is negligible in this range). Outside this
range the flux calibration can introduce significant structures
as the throughput of the whole system drops very rapidly, thus
even very small inaccuracies in the wavelength calibration of
the object spectra result in significant over or underestimation
of the physical spectra. It is also important to point out that
no attempt was made to correct for telluric absorption: For the
vast majority of our program objects, telluric absorption
completely eliminates the signal (between 7600-7630{\AA},
7170-7350{\AA} and 6868-6890{\AA}), thus a correction is not
practical. For the few brighter objects, this correction would
have been possible, but was deemed unnecessary for
identification of the sources.

\subsection{\it Absolute Calibration -- Estimating the Slitloss\label{abscal}}

The purpose of our observations was to identify as many X-ray sources as
possible with the telescope time available. Therefore, no effort was made
to collect data necessary for {\em absolute} calibration of the spectra
measured. Even though we did {\em not} use an elaborate program designed
for spectrophotometry, we can still estimate the accuracy of our
derived fluxes.

The most important effect to be considered is slit-loss. To maximize the
S/N of the data, we tried to match the slit width to the
expected seeing of the observations. Therefore, a significant fraction
of the light from the object was excluded, but this was more than
balanced by the large reduction in the sky background, and 
thus the increase in the S/N of the source.

We can easily estimate the effect of slit-losses from the 
high accuracy broad-band photometry. Using the
flux calibrated spectra, we can directly calculate the AB-magnitude
of the object in any filter:
\begin{equation}
m_{AB}=-2.5 \log{\int d(\log \nu) f_\nu S_\nu\over \int d(\log \nu)S_\nu}-48.60,
\label{ABmag}
\end{equation}
where $f_\nu$ is the energy flux per unit frequency, $S_\nu$ is the
overall throughput of the system (telescope and instrument) in arbitrary
units.

The first step is to select the filter curve to use,
$S_\nu$. In practice, no system can replicate the canonical
Cousins-Johnson filter curves {\em exactly}. Even a perfect
filter response curve would be distorted by the non flatness of the CCD
detector. In many cases a slight deviation from this filter response
curve is acceptable, assuming that the spectrum of the object is
smooth and the slope is not very different from the slope of
Vega. Unfortunately, these assumptions do not hold for most of
our objects as a significant fraction of the flux is in
very sharp features. Therefore, we have to use the effective
filter curve of {\em our} system used to derive the broad-band
magnitudes. Fortunately,
the Bessel filter set used by ESO is a
sufficiently good approximation of the Cousins-Johnson filters and
the quantum efficiency curve of the FORS detectors being relatively
flat, this correction would only amount to a fraction of a
percent and can be safely ignored. Therefore, we used the
published ESO filter response curves folded over the quantum efficiency
of the detectors as system throughput, $S_\nu$.
To convert to Vega magnitudes, we calculated the AB magnitude of Vega from
its spectra \citep{fukugita1996}. The resulting slitlosses are presented
in Table \ref{tblobs} for point sources for each mask. We also checked if
the slitloss depends on wavelength (by comparing different broad-band
magnitudes) but found no significant effect.

\subsection{\it Reddening Correction\label{deredden}}

To calculate the effect of reddening due to our Galaxy on the
spectra, we used the 100$\mu m$ maps \citep{schlegel1998}.
In the direction of the CDFS, $l=223.5\degr$, $b=-54.4\degr$,
the color excess is $E$(B$-$V)$\approx0.008$. Assuming the canonical value,
$R_{\rm V}=3.1$ for the ratio of extinction in the V-band to the
color excess \citep{cardelli1989},
the extintion is $A$(U)$\approx 0.04$
and $A$(I)$\approx 0.01$. As this extinction is heavily dependent
on the choice of $R_{\rm V}$, this correction was {\em not} applied to
the data, introducing an artificial tilt in all spectra on the
order of a few percent.

The AGN line strengths were not corrected for
absorption lines from the {\em host} galaxy \citep{ho1993}
as the S/N of our faint spectroscopic sample is too low.
Consequently, the very few optical identifications based on line 
ratios are possibly affected by this effect.

Correcting for reddening by the AGN host galaxy would require to estimate 
the extinction using the X-ray spectral information. This correction is 
deferred to a later paper concentrating on X-ray spectral analysis of the 
CDFS sources based on the Chandra and XMM-Newton data.

\section{REDSHIFT DETERMINATION AND THE SPECTROSCOPIC SAMPLE\label{zestimate}}

\subsection{\it Redshift and Luminosity Determination}

The first step toward the classification of the spectroscopically  
observed sources was their {\em redshift} determination.
In the vast majority of the cases this was done through the
identification of prominent features, typically the 4000{\AA}
break and the Ca\,{\sc ii} H and K absorption, Balmer lines  or emission
lines (e.g. Ly-$\alpha$, C\,{\sc iv}, C\,{\sc iii}], Mg\,{\sc ii},
[O\,{\sc ii}], etc.). In case of
prominent emission lines, the
wavelength ratio of the line centers was used to identify these
features. In cases of single emission line objects with no
additional feature, this line was usually identified as either [O\,{\sc ii}] 
or Ly-$\alpha$, depending on the continuum spectral shape. 
Naturally, these are not secure
classifications, the quoted redshifts should only be used as an
educated guess to optimize follow-up observations.

The redshift identifications are summarized in Table \ref{tblspec}.
The `No' column refers to our internal id of the {\em X-ray} source -- this
is the {\em unique detection ID (XID)} in the published catalog \citep{giacconi2002}. In cases
of multiple counterparts, a letter is appended to this number to 
distinguish between the optical candidates. Extended X-ray objects are marked
with a star.
When an object was observed 
repeatedly, multiple entries are given in the table. 
Altogether 249 X-ray sources were observed, of which one point source 
belongs to the small additional sample given in Table \ref{newsrc}, and 
15 are extended X-ray sources. In 17 cases, 
the slit was centered on the X-ray position for the search of strong, 
narrow emission lines although no counterpart was detected in the R-band.

The {\em mask} column is our internal name used to identify the
set of observations used for individual objects. Multiple
mask names indicate that during the observations some slits were
reconfigured, but the slit used for the
object was identical during the set of observations.  The
relevant observing conditions and configuration can be found in
Table \ref{tblobs} (exposure time, slit width, seeing, etc.).

The two position columns (right ascension and declination) give the coordinates
of the {\em optical} object, {\em not} those of the X-ray source.
Astrometry is based on the USNO \citep{usno} reference frame, just like
the X-ray positions in \citet{giacconi2002} and the astrometric accuracy
is better than 0.2\arcsec.

Whenever available, we also provide
broadband optical information, an R-band magnitude and 
R$-$K color (both in Vega magnitudes). If no R-band magnitude is given,
it implies that our FORS imaging data is not deep enough to measure the
magnitude of the object (all program objects are covered by the FORS
R-band survey). The lack of R$-$K color can be due
to our limited near-infrared coverage ({\em NA} entries).

Assuming (throughout this paper) an $\Omega_m=0.3$, $\Omega_\Lambda=0.7$
universe and 
$H_0=70$ km s$^{-1}$ Mpc$^{-1}$ \citep{spergel2003}
the total X-ray intrinsic luminosity of 
the object, $L_{\rm X}$, in erg s$^{-1}$, is \citep{carroll1992}
\begin{equation}
L_{\rm X}(f_{\rm X},z)=4\pi f_X
\left(
{c(1+z)\over H_0\sqrt{\vert\Omega_k\vert}}\sinn\left(\sqrt{\vert\Omega_k\vert}
\int\limits_0^z \left(
(1+\zeta)^2(1+\Omega_m\zeta)-\zeta(2+\zeta)\Omega_\Lambda
\right)^{-1/2}d\zeta\right)\right)^2
\end{equation}
where $f_{X,tot}$ is the {\em observed} X-ray flux in the 0.5-10 keV
band. The $\sinn(x)$ function is $\sin(x)$ for $\Omega_k<0$,
$\sinh(x)$ for $\Omega_k<0$ and simply $x$ for $\Omega_k=0$, where
$\Omega_k\equiv 1-\Omega_m-\Omega_\Lambda$. In case of $\Omega_k=0$,
the two $\sqrt{\vert\Omega_k\vert}$ terms disappear.
This flux being the {\em observed}
X-ray flux, it should be interpreted as a lower limit
for the {\em intrinsic} X-ray flux of the object due to potentially
strong obscuration of the source.

If we assume that $\Omega_k=0$ (i.e. $\Omega_m+\Omega_\Lambda\equiv1$), we
can rewrite this equation as
\begin{equation}
L_{\rm X}(f_{\rm X},z)=4\pi f_X
\left(
{c(1+z)\over H_0}\left(
\int\limits_0^z \left(
(1+\zeta)^3\Omega_m+\Omega_\Lambda
\right)^{-1/2}d\zeta\right)\right)^2
\end{equation}

Introducing $x=(1+\zeta)(\Omega_m/\Omega_\Lambda)^{1/3}$, this simplifies to
\begin{equation}
L_{\rm X}(f_{\rm X},z)=
{4\pi f_X c^2(1+z)^2\over H_0^2\Omega_m^{2/3}\Omega_\Lambda^{1/3}}\left(
\int\limits_{(\Omega_m/\Omega_\Lambda)^{1/3}}^{(1+z)(\Omega_m/\Omega_\Lambda)^{1/3}}
{dx\over\sqrt{x^3+1}}
\right)^2
\end{equation}

The above integral can be evaluated in 
terms of incomplete elliptical integrals of the first kind.
We also give below an analytical fit for flat cosmologies
\citep{pen1999} with $\Omega_m=0.3$:
\begin{equation}
d_L={c(1+z)\over H_0}\left(3.308-3.651
	\left(0.207 + 
              0.446(1 + z) + 
              0.757(1 + z)^2 - 
              0.204(1 + z)^3 + 
                   (1 + z)^4
	\right)^{-1/8}\right)
\end{equation}

\begin{figure}
\plotone{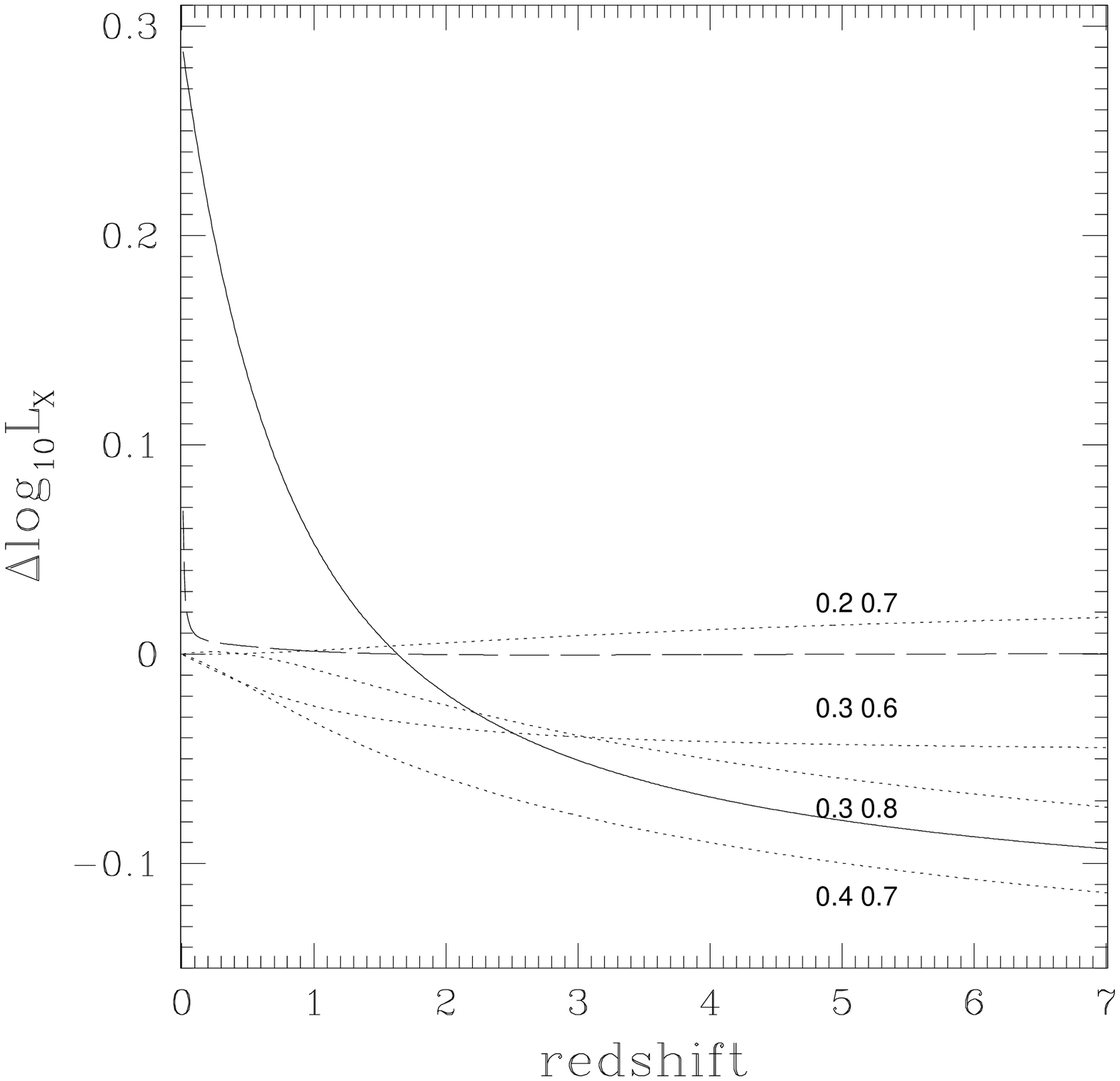}
\caption{Comparing calculated $L_X$ values in different cosmologies.
Plotted are the changes as a function of redshift, compared to 
the current $\Omega_m=0.3$, $\Omega_\Lambda=0.7$, $H_0=70 km\ s^{-1}\ Mpc^{-1}$
cosmology. Dotted lines are slight variations on the two basic parameters,
without changing the Hubble constant. The solid line is the now obsolete
$\Omega_m=1$, $\Omega_\Lambda=0$, $H_0=50 km\ s^{-1}\ Mpc^{-1}$
cosmology. The result of an approximate formula \citep{pen1999} for
$\Omega_m=0.3$, $\Omega_\Lambda=0.7$, $H_0=70 km\ s^{-1}\ Mpc^{-1}$
is also shown (dashed line).\label{lxcomp}}.
\end{figure}

In Figure \ref{lxcomp} we show the correction to the calculated X-ray 
luminosity for slightly different cosmologies. We also show the 
difference between luminosities calculated using current cosmological 
parameters and the (now obsolete)
$\Omega_m=1$, $\Omega_\Lambda=0$, $H_0=50$ km s$^{-1}$ Mpc$^{-1}$ 
cosmology,
\begin{equation}
L_{\rm X}=4\pi f_{X,tot}\left({2 c \over
H_0}\left(1+z-\sqrt{1+z}\right)\right)^2
\approx
1.72\times 10^{58} \cm^2 f_{X,tot}\left(1+z-\sqrt{1+z}\right)^2,
\end{equation}

The $HR$ column contains the already published \citep{giacconi2002}
hardness ratios for each object, $HR=(H-S)/(H+S)$, where $H$ and $S$
are the net count rates in the hard (2-10 keV) and soft (0.5-2 keV)
band, respectively. It is important to point out that the hardness ratio
is defined in {\em instrument} counts (for Chandra ACIS-I), thus
for different X-ray telescopes or instruments, 
it should be converted using their specific energy conversion factors.

The $z$ column gives our best redshift estimate. The selected low spectral 
resolution leads to an uncertainty in the redshift determination of 
$\pm0.005$. For broad emission line objects the uncertainty is 
significantly higher. The quoted redshift value {\em always} refers to 
the particular observation of the object, thus, there can be slight 
discrepancies between observations of the same object or missing redshift 
values for some masks.

\subsection{\it The Optical and X-ray Classification}

The classical/optical and X-ray classifications of the objects
 are discussed in details in Section \ref{classes} and given in 
 Table \ref{tblspec}. 

Based on purely the optical spectra, we define the following {\em optical}
object classes:

\begin{itemize}
\item {\em BLAGN}: Objects with emission lines broader than 2000 km s$^{-1}$.
This classification implies an optical type-1 AGN or QSO, as discussed in
Section \ref{classes}.
\item {\em HEX}: Object with unresolved emission lines {\em and} exhibiting
high ionization lines or emission line ratios indicating AGN activity.
These objects are dominanly optical type-2 AGNs or QSOs, but in a few cases the
optical type-1/2 distinction is not possible based on the data.
\item {\em LEX}: Objects with unresolved emission lines consistent with an
H\,{\sc ii} region-type spectra. These objects would be classified as
normal galaxies based on the optical data alone as the presence of the
AGN can not be established.
\item {\em ABS}: a typical galaxy spectrum showing only absorption lines.
\item {\em star}: a stellar spectrum.
\end{itemize}

Our main classification is solely based on the observed X-ray properties 
($L_{\rm X}$ and $HR$) of the sources and is summarized below. The 
type-1 AGN/QSO are soft X-ray sources, while the type-2 AGN/QSO are hard,
absorbed X-ray sources \citep[for the relationship between hardness 
ratio and absorption see e.g.][]{mainieri2002}. The AGN and QSO classes 
cover different ranges of X-ray luminosities. 

\begin{itemize}
\item 
{\em QSO-1}: $L_{\rm X}$(0.5-10 keV)$\geq10^{44}~\erg~\s^{-1}$ and 
$HR\leq-0.2$.
\item 
{\em AGN-1}: $10^{42}\leq L_{\rm X}$(0.5-10 keV)$<10^{44}~\erg~\s^{-1}$ and 
$HR\leq-0.2$.
\item 
{\em QSO-2}: $L_{\rm X}$(0.5-10 keV)$\geq10^{44}~\erg~\s^{-1}$  and $HR>-0.2$.
\item 
{\em AGN-2}: $10^{41}\leq L_{\rm X}$(0.5-10 keV)$<10^{44}~\erg~\s^{-1}$ 
(lower limit smaller than for the AGN-1 population to account for 
substantial absorption)  and $HR>-0.2$.
\item 
{\em gal}: $L_{\rm X}$(0.5-10 keV)$<10^{42}~\erg~\s^{-1}$ and $HR<-0.2$.  
\item 
{\em star}: this class is defined from the optical spectra and/or 
proper motions.
\end{itemize}

Throughout the paper we call X-ray type-1/2 AGNs and QSOs together X-ray
type-1/2 {\em objects}.

\subsection{\it The Spectroscopic Sample}

The {\em quality} flag, $Q$, indicates the reliablity of the 
redshift determination. $Q=2.0$ indicates a reliable redshift 
determination, a value of 0.0 indicates no success. $Q=1.0$ indicates 
that we clearly detect {\em some} feature (typically a single narrow 
emission line) in the spectrum that cannot be identified securely. 
In a few cases, $Q=0.5$ is used when there is a hint of some spectral feature. 
This quality flag {\em only} refers to the reliability of the 
spectroscopic classification. The identification of the X-ray sources 
is unambiguous for single counterparts in the X-ray error circles, 
and for cases with reliable redshift identification we then use a 
$Q=2.0+$ quality flag. X-ray sources with multiple counterparts are 
discussed below. The $Q=2.0+$ objects define our spectroscopically identified 
X-ray sample.  

Finally, the {\em comments} column contains additional information relevant
to the particular observation. The most common ones are a limited wavelength 
coverage (the full wavelength range is not available due to the positioning 
of the slit) and the detection of high ionization lines. We also include
information necessary to apply the stricy Seyfert definition
\citep{khachikian1974} to optically classify our objects.

Among the X-ray sources with multiple optical counterparts, the 
identification is considered as highly reliable in the following 13 cases:
X-ray type-1 QSO/AGN (XID 30, 101), X-ray type-2 QSO/AGN (XID 56, 201, 263), 
and interacting galaxy pairs (XID 98, 138, 580). 
Four of the remaining X-ray sources 
(XID 553, 567, 582, 620) are the brightest objects in the X-ray 
error circles, only detected in the soft band and at moderate redshift
with H\,{\sc ii} region-type spectra, thus most likely the X-ray 
counterparts. The last source (XID 189) is detected in the hard band only, 
has R$-$K $>$ 5 and is the brigthest, best centered counterpart 
(the additional counterparts are very faint, R $>$ 25).
These objects are included in the spectroscopically identified 
X-ray sample.

As the CDFS has been observed by various teams and covered by
wide surveys (2dF and Tycho), we could include in our sample
additional spectroscopic redshifts.  Four sources (XID 39, 95,
103, 116) were in fact already published in the CDFS 130 ksec
paper \citep{giacconi2001}. Eight objects (XID 33, 38, 149, 171,
204, 526, 563, 600) are covered by the K20 survey
\citep[][Cimatti, private communications]{daddi2003} and three
(XID 90, 92 and 647) by the COMBO-17 survey (Wolf, private
communication).  Three sources have optically bright, low $z$  
counterparts in the 2dFGRS \citep{colless2001}. One ot them, XID 84 
(TGS243Z005), has a soft X-ray spectrum and appears to be a normal 
X-ray galaxy. The other two sources, XID 247 (TGS243Z011) and  
XID 514 (TGS243Z010),  have hard spectra, luminosities  
$L_{\rm X}$(0.5-10 keV) $\sim10^{40}~\erg~\s^{-1}$ and are off-centred
within the parent galaxy: these properties are similar to those 
of the brigther, ultraluminous compact X-ray sources (ULXs) detected 
in nearby spiral galaxies \citep{makishima2000}.
Finally, one source (XID 549), optically very bright, is
identified with an object (TYC 6453-888-1) from the Tycho
Reference Catalog \citep{hog1998}, that has a clear proper
motion (7.6$\pm$2.1 and 15.7$\pm$1.7 mas/yr in right ascension
and declination, respectively) and is thus a star in the
Milkyway.

The spectroscopically identified X-ray sample comprises 137 sources of which
15 are fainter than R = 24.0. Among the brighter objects, there are seven 
extended X-ray sources at $ 0.6 < z < 1$
\citep[XID 132, 138, 249, 560, 566, 594, 645:][]{giacconi2002}. 
In addition, there are 24 X-ray sources with secure identifications but 
only tentative redshifts ($Q=0.5$ and 1.0 cases). 

\section{FINDING CHARTS AND SPECTRA}

Figure \ref{spectra} gives the finding charts and {\em all} the spectra
obtained for each of our program objects. 

Finding charts are $20\arcsec\times20\arcsec$ in size, centered on the X-ray 
position with its $2\sigma$  position error circle.
Individual contrast levels are chosen in each case to give as much
information as possible. If multiple optical counterparts are present
in (or around) the error circle, they are marked and labeled. The underlying
optical images are our $R$-band FORS images.

Next to the finding charts, we show the associated spectra. For cases of
repeated observations, all the data are shown. In the
plots, we mark the features used for the redshift determination.

In cases of marginal line detections, we also examined the sky-subtracted
coadded two-dimensional frames to confirm/infirm the presence of
the feature. These images are not included in this paper, but are
available through our web-site, \url{http://www.mpe.mpg.de/CDFS}.

\section{FIELD SAMPLE\label{fieldsample}}

During our survey, we also collected a large number of field object
spectra. These were objects either accidentally covered by some
of our slits, or observed in slits that could 
not be placed on X-ray counterpart candidates due to geometrical constraints, 
or stars used to align the MXU masks. Consequently, this sample is
not representative of the field population.

The results of these observations are summarized in Table \ref{tblfield}.
We give the object position, mask name, $R$-band magnitude (when available),
redshift, redshift quality flag, very crude
classification and optional comments relevant to the observation. The
full dataset (spectra and finding charts) is available on our web site
(\url{http://www.mpe.mpg.de/CDFS}).

\section{X-RAY VERSUS OPTICAL CLASSIFICATION}\label{classes}

Seyfert galaxies \citep{seyfert1943} were originally
defined \citep{khachikian1974} as a recognizable galaxy (on {\em
Sky Survey} prints) that have broad ($>500$ km s$^{-1}$) emission lines 
arising in a bright, semi-stellar nucleus. Seyfert galaxies were 
subdivided into class 1 and class 2, depending on the width of the 
Balmer lines, compared to that of forbidden lines. 
For line widths $200<FWHM<500$ km s$^{-1}$,
additional criteria were applied, based on emission line ratios
\citep[e.g.][]{osterbrock1989} to establish the Seyfert nature
of a galaxy.  With the emergence of the unified AGN model
\citep{antonucci1985}, it is now widely accepted that these two
Seyfert classes are not distinct, but form a continuous
distribution between the two extremes and a large number of
intermediate classes were introduced since the original definition.  
Applying these classical definitions poses
very serious problems for the study of faint X-ray sources.

At high redshifts, the Balmer lines are no longer in the optical 
range: the two strongest Balmer lines, H$\alpha$ and H$\beta$, 
are redwards of 8500{\AA} for $z>0.3$ and 0.75, respectively. 
This problem was overcome by extending the original definition
to {\em permitted} lines in the UV range from 
Mg\,{\sc ii}$\lambda\lambda$2796,2803 to Ly$\alpha$
\citep[see e.g.][]{schmidt1998}, thus 
allowing an optical classification of objects up to $z\sim6.5$.
Another difficulty stems from the fact that most of
the objects associated with faint X-ray sources are at intermediate
redshifts and, thus, are comparable in size to the seeing
achievable with ground-based optical telescopes. As a
consequence, we can only study the {\em integrated} emission
from these objects, as opposed to {\em nuclear} emission from
local Seyfert galaxies. Consequently, the nuclear emission can be
`hidden' in the stellar light coming from the host galaxy. The
study of local Seyfert galaxies confirms that about 60\% of the
local Seyfert type-2 galaxies would {\em not} be classified as
Seyfert-2, if only the {\em total} emission were available
\citep{moran2002}.

Moreover, an obvious challenge in applying the classical Seyfert
definition for faint objects is merely to recognize that they are 
AGNs. The main optical classes introduced in this paper for 
extragalactic sources are: 1) BLAGN -- $FWHM$(permitted lines)$>2000$ km 
s$^{-1}$, 2) HEX -- unresolved emission features 
but presence of high excitation lines not found in H\,{\sc ii} regions 
(e.g. [Ne\,{\sc v}]$\lambda$3425, He\,{\sc ii}$\lambda$1640), suggesting 
AGNs of the optical type-2 class, 3) LEX -- H\,{\sc ii} region-type 
spectrum, 4) ABS -- typical galaxy absorption line spectrum. For the 130 
extragalactic X-ray sources with secure redshift identification, there are 
32 BLAGN, 24 HEX, 54 LEX and 21 ABS sources, thus 57\% LEX+ABS objects. 
But among the latter (optically dull), it should be noted that there is a large 
number of luminous X-ray sources. 

To overcome the limitations of the classical/optical definition
of AGN, we follow the unified AGN model introduced by
\citet{antonucci1985} and classify an object as an AGN
{\em if it has (nuclear) emission stronger than expected from
stellar processes in normal galaxies}. This emission is likely
to be produced by strong accretion onto supermassive objects, most
probably black holes.  A clear signature of the presence of
this accretion is a high X-ray luminosity.

An X-ray classification requires first to introduce a conservative lower 
limit on the (unabsorbed) absolute X-ray luminosity of AGNs. 
Local, well studied starburst galaxies have X-ray luminosities in the 
0.5-10 keV band typically below 10$^{42}$ erg s$^{-1}$ 
\citep{rosati2002,alexander2002}. Thermal haloes of galaxies and 
intragroup/cluster gas can have higher X-ray luminosities but, in Chandra 
data, they are spatially resolved  and detected only in the soft band 
thus, at intermediate redshifts, they become fainter than 
$10^{42}$ erg s$^{-1}$ in the 0.5-10 keV band. 
Accordingly, objects with $L_{\rm X}$(0.5-10 keV) $\geq 10^{42}$ erg s$^{-1}$,
should be classified AGNs. There are 20[20] HEX, 31[53] LEX and 9[12] ABS 
(excluding XID 645 which only shows extended X-ray emission) high  $L_{\rm X}$ 
sources with secure[secure+tentative]  redshift identification respectively. 
Thus the optical classification completely fails to identify as AGN 42\% 
(LEX+ABS fraction) of the luminous X-ray sources (96), or altogether 54\% if we 
include the tentative redshift identifications (120). In these objects, optical
extinction of the nuclear component  by dust can be very high, and/or the 
host galaxy can outshine the central AGN \citep{lehmann2000,lehmann2001}. 

The X-ray luminosity is also used to separate the sources of the AGN class, 
$10^{42} \leq L_{\rm X}$(0.5-10 keV) $< 10^{44}$ erg s$^{-1}$, from those 
of the QSO class, $L_{\rm X}$(0.5-10 keV) $\geq 10^{44}$ erg s$^{-1}$.

\begin{figure}
\plotone{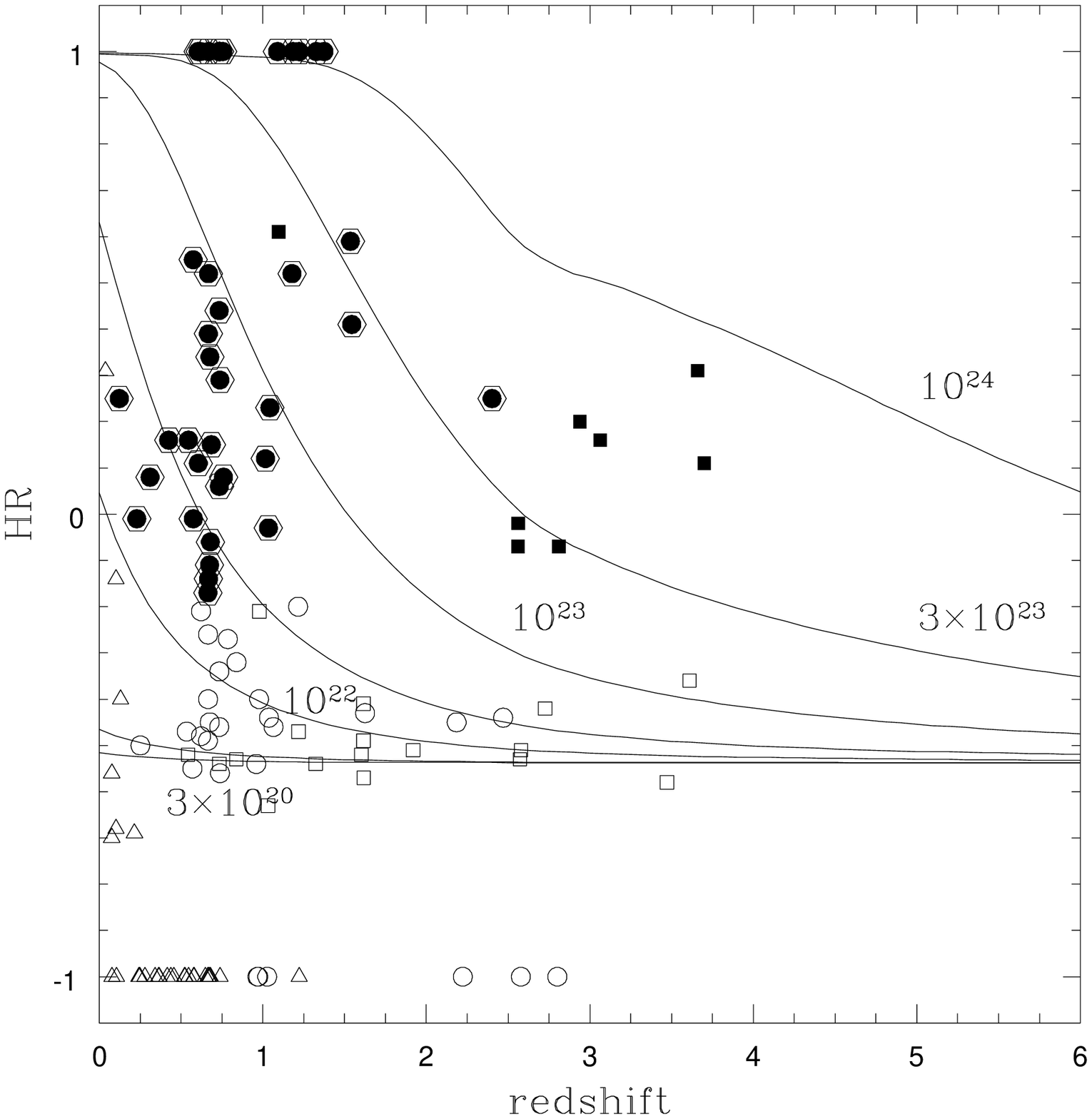}

\caption{Hardness ratio versus redhsift, assuming a typical absorbed
AGN spectrum with $\Gamma=2$ and different absorptions, expressed
as $\log(N_H)$. Identified CDFS objects are also plotted (see Figure 
\ref{fxR} for symbols). HEX objects (likely optical type-2 AGNs) are
additionaly marked with a hexagon.
\label{HR_z_sim}}

\end{figure}

Secondly, following the unified AGN model, we can also define two AGN classes
by using the hardness ratio, a parameter sensitive to X-ray absorption
which can be measured even for faint objects. In Figure
\ref{HR_z_sim}, we give the expected hardness ratios for AGNs with power 
law X-ray spectra, selecting a photon index $\Gamma$=2 and different
absorption levels. Unabsorbed sources have $HR\approx-0.5$, 
independent of $z$. This is indeed the case for all the BLAGNs; their hardness 
ratios are in the range $-1.0\leq HR \leq-0.2$, except for one BAL QSO.  
The scatter is easily explained by introducing
different slopes for the X-ray spectra, together with statistical errors
associated with low number counts in the X-ray bands. 
The harder spectra ($HR>-0.2$) are fully  consistent with {\em absorbed} 
power law spectra. Significant intrinsic absorption, $10^{21.5} <$
$N_{\rm H}$ $\lesssim 10^{23.5}$~cm$^{-2}$,  has indeed already been found 
for the type-2 AGN population \citep{mainieri2002,barger2002}.
Figure \ref{HR_z_sim} shows that, assuming $\Gamma=2$, intrinsic 
absorption ($HR>-0.2$) can be detected up to $z$ = 0.25, 2.1 and 3.9
for $N_{\rm H}=10^{22}, 10^{23}$ and $3\times10^{23}$ cm$^{-2}$, 
respectively. Thus the hardness ratio can be used to separate the unabsorbed 
sources, X-ray type-1: $HR\leq-0.2$, from the absorbed ones, 
X-ray type-2: $HR>-0.2$. 
Indeed, in the Chandra and XMM-Newton deep surveys, most of the harder
X-ray sources are optical type-2 AGN with an increasing fraction of
absorption at decreasing X-ray flux \citep{barger2001a,
barger2001b, hasinger2001, rosati2002, mainieri2002}.  Among
this class of objects, there are a few bright type-2 QSOs but
the majority of the sources are type-2 AGNs at $z \lesssim 1$
\citep[see e.g.][]{hasinger2002}. It should be noted that an X-ray 
classification based on the hardness ratio  might be misleading for some 
high-redshift objects: an increasing absorption makes the sources harder, 
while a higher redshift makes them softer. Consequently, some high-redshift 
absorbed/type-2 sources may be mistakenly identified as type-1, but
not the other way around. 

A consistent X-ray classification should use the intrinsic
luminosity.  As mentioned above, hard sources ($HR>-0.2$) at
$z\sim0.25$ have absorbing column densities $N_{\rm
H}\geq10^{22}$ cm$^{-2}$ and thus their de-absorbed flux in the
observed 0.5-10 keV band is at least 5 times larger than the
observed flux.  Consequently, {\em hard} objects ($HR>-0.2$)
with lower luminosities (($10^{41} < L_{\rm X}$(0.5-10 keV) $<
10^{42}$ erg s$^{-1}$) can be classified as X-ray type-2
AGN.  Four additional objects (XID 55, 525, 538, 598) are thus
classified as low $L_{\rm X}$ X-ray type-2 AGNs.

In Figure \ref{hrlx}, we show the hardness ratio versus the observed 
X-ray luminosity for all the sources with secure redshift, for both  
the optical classification (left panel) and the X-ray one (right panel).
No source with a very high X-ray luminosity is present in this diagram: 
this is, at least in part, a selection effect of pencil beam surveys. 
We now compare the optical and the X-ray classifications.

\begin{itemize}
\item
Of the 32 BLAGNs in our sample, all are X-ray type-1 objects, except 
the BAL QSO (XID 62, $HR=-0.07$) which is an X-ray type-2 QSO.

\item
Among the HEX population (24 objects), there are 16 X-ray type-2 AGNs/QSOs (including 
one low $L_{\rm X}$ source: XID 55) for which X-ray absorption is indeed 
associated with optical obscuration. There are eight X-ray type-1 AGNs/QSOs or
galaxies,
of which four at $z\geq1.6$ (XID 31, 117, 563, 901) with permitted emission 
lines no broader than $\sim1500$ km s$^{-1}$. These four sources may be 
partly absorbed ($N_{\rm H}=10^{22}$-$10^{23}$ cm$^{-2}$), thus 
misclassified as X-ray type-1: the presence of probable X-ray absorption 
should  be confirmed by X-ray spectral analysis, whenever possible. 
In the spectrum of XID 34a, we do not detect permitted lines.
The remaining three HEX objects are X-ray galaxies (XID 98a, 175b, 580a), 
two being  members of  interacting pairs, and all have $HR=-1.0$. They 
could be either low $L_{\rm X}$ type-1 AGN or, in the case of the interacting 
pairs, shocks might be at the origin of the [Ne\,{\sc v}] emission.\\
There is a high fraction, 42\%, of $z>2$ 
sources among the HEX class as compared to  25\% in the BLAGN class.
 
\item
The LEX population comprises 54 sources with secure redhift identification,
of which 9 and 24 X-ray type-1 and type-2 (including two low $L_{\rm X}$ 
sources: XID 525, 538) AGNs/QSOs, respectively. The optical
classification thus fails to identify as AGN  61\% of this population. 
Among the remaining sources, there are 21 X-ray galaxies including one 
ULX (XID 247) at $z=0.038$ with a hard spectrum, $HR=0.31$ 
(see Section \ref{zestimate}). In the LEX class, there are only two 
high luminosity sources of the X-ray QSO class and no objects 
at high redshift ($z>1.5$).  
A few X-ray type-2 AGNs might be of the HEX class but, due to the 
low S/N ($<5$ per resolution element) of their optical spectra, 
high excitation lines could be below the detection threshold. \\
  For most X-ray type-2 AGNs,  both the broad (BLR) and  narrow 
(NLR) emission line regions could be obscured by dust absorption.
Alternatively,  the obscuring region may fully cover the central UV 
source and the BLR, preventing photo-ionization of external regions 
thus the existence of a NLR.
  The AGN nature of all the X-ray type-1 sources ($0.53\leq z\leq 1.03$)
is difficult to ascertain from the optical data alone as the 
H$\alpha$ line is outside the observing range for redshift higher than
0.4. For six of them, the 
expected Mg\,{\sc ii} emission line is within the observed range (i.e.
$0.4<z<2.2$) and 
away from strong sky lines, but the S/N is not high enough to detect a 
weak broad line; in one source (XID 138), a broad  Mg\,{\sc ii} emission 
line may be present, although at a low significance level. For 
comparison, Mg\,{\sc ii} is often seen in absorption or with a P Cygni 
profile in star-forming galaxies \citep{kinney1993}.

\item
There are 21 sources in the ABS class of which one and 9 X-ray type-1 
and type-2 (including one low $L_{\rm X}$ source: XID 598) AGNs. 
Thus 48\% of the AGN population in the ABS class is missed by the 
optical classification. All the sources of the X-ray AGN class are 
at $z<1.2$.   The 11 X-ray galaxies are all at $z<0.8$ and have
soft spectra ($HR<-0.7$) except one object (XID 514), a ULX at 
$z=0.103$ with  $HR=-0.14$ (see Section \ref{zestimate}).

\end{itemize}

The comparision of the two classification schemes are summarised in Table
\ref{comptab}.

The proposed X-ray classification is more successful than the 
classical/optical one in revealing the presence of black hole activity,
whatever the amount of dust obscuration from the central and/or 
external parts of the nuclear region. Thus, we use this classification 
throughout the paper unless otherwise stated. For comparison, we also 
give the optical classification in Table \ref{tblspec}. The latter 
may be more appropriate in studies that aim to extrapolate the 
classical Seyfert definition to faint AGNs. 

It should be noted that using the X-ray classification is mandatory to 
properly identify the X-ray normal galaxies among the LEX+ABS optical 
class. This population provides another means to derive the star formation 
history of the universe, in addition to the methods using radio or optical 
data.


\section{CLUSTERS AND EXTENDED SOURCES}

Of the 19 extended sources detected in the CDFS \citep{giacconi2002},
15 were observed in our survey. In 5 cases (XID 37, 147, 522, 527 and
581) no spectroscopic identification was possible and in one case (XID
132) only a low quality identification was obtained. In two cases
(XID 116 and 514) the diffuse X-ray emission could be ascribed to 
thermal halos of nearby galaxies. 
In general, most
of the remaining extended sources span the regime of galaxy groups (with
luminosities of a few$\times 10^{42}$ erg s$^{-1}$) down to X-ray
luminosities typical of thermal halos around single early-type
galaxies. In some cases, either the hardness ratio or the optical
identification suggest the coexistence of a thermal halo with an AGN
component (e.g. XID 138). In Figures \ref{cdf138}-\ref{cdf645} we 
show K-band images of the identifield clusters/groups with overlaid Chandra contours
(2.5,3,4,5,7,10 $\sigma$ above the local background) in the [0.5-2]
keV band. We also mark objects with concordant redshifts (as listed in
Table 5).

Specifically, XID 566,594,645 are ordinary groups showing
however a range of surface brightness profiles (see Figs.
\ref{cdf566},\ref{cdf594},\ref{cdf645}).
XID 566 and 594 belong to the large scale structure at $z\simeq
0.73$ (see below). XID 249, for which we have two concordant
redshifts with $<z>=0.964$, is clearly extended with a harder
component (Fig. \ref{cdf249}).  XID 138 was identified as
a close pair of AGN at $z=0.97$, surrounded by a soft halo.  In
two cases, XID 511 and 560, we identified only one galaxy per
source, making it difficult to ascertain the existence of a
group.

\section{REDSHIFT DISTRIBUTION\label{reddis}}
 
The spectroscopically identified CDFS sample comprises 135 
X-ray sources, including five stars. Reliable redshifts can be obtained 
typically for objects with R $<25.5$, however, some incompleteness already
sets in around R $\sim23$. For the R$<$24 sample (199 objects), 120 
(including five stars) of the 159  observed X-ray sources have been 
spectroscopically identified, thus a success rate of 75\% and
a completenes of 60\%. 
The sources with inconclusive redshift identification cover a wide range of 
hardness ratios. In Figure \ref{position}, we show the spatial distribution 
of the sources with spectroscopic observations as well as those not observed. 
The latter lie predominantly in some of the outermost parts of the CDFS.

\begin{figure}
\plotone{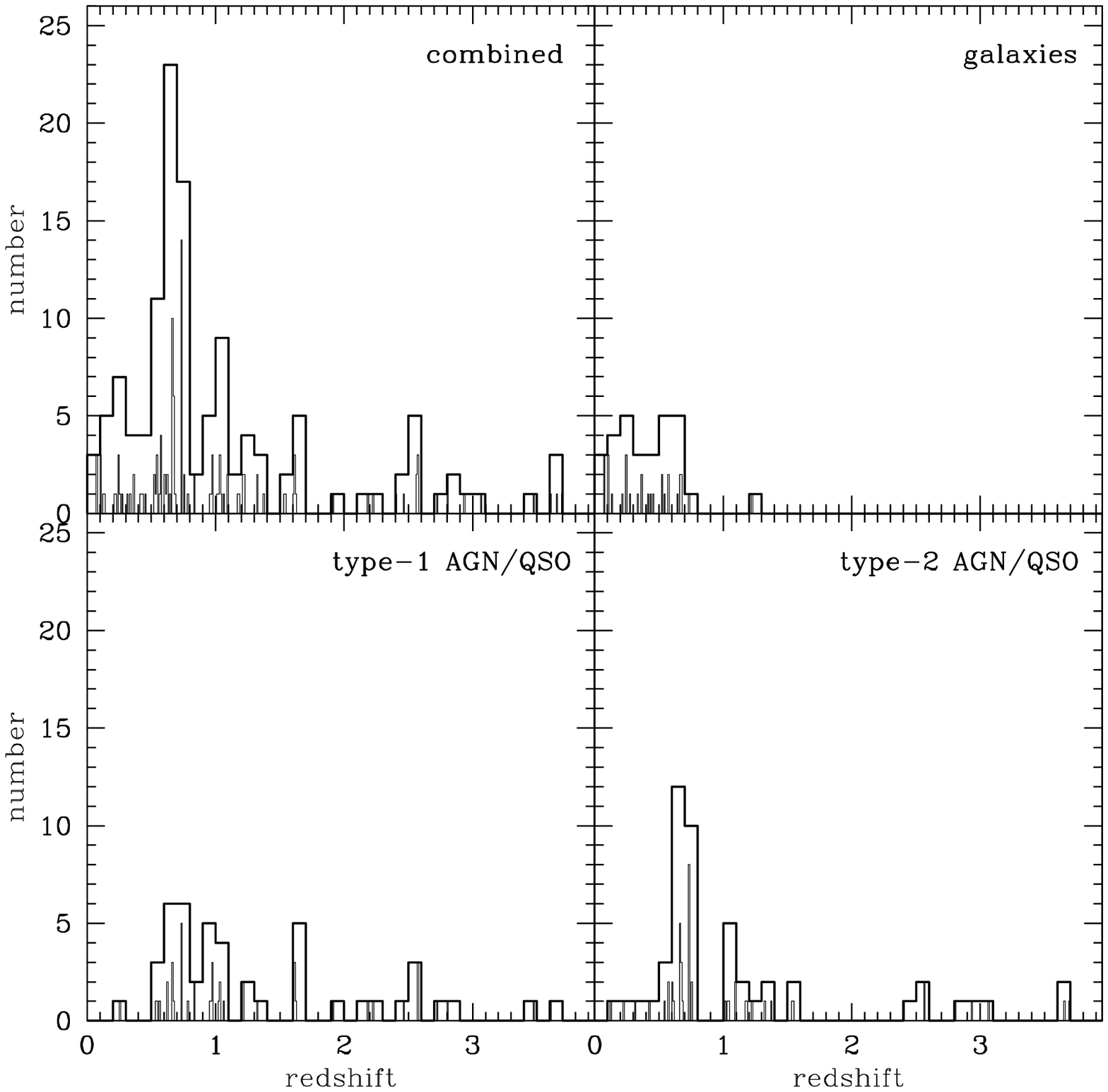}

\caption{Redshift distribution of the various classes of X-ray sources
as well as their cumulative distribution. Object XID 138a and 138b are
counted as two separate objects. The 4 extended objects (249, 566, 594
and 645) are not included. Two bin sizes
($\Delta z =$ 0.01 and 0.1) are selected to show the excess of objects 
around  $z=0.674$ and 0.734. 
\label{zdist}}

\end{figure}

The histogram of the redshift distribution of the X-ray sources is shown 
in Figure \ref{zdist}. A preliminary version of this diagram was given by 
\citet{hasinger2002}. There is an excess of objects in two redshift bins,
revealing large-scale structures of X-ray sources\citep{gilli2003}, similar 
to that found in the 
CDFN \citep{barger2002}. These redshift spikes are populated by X-ray 
type-1 and type-2 AGNs as well as a few X-ray galaxies. 
There are 18 X-ray sources within 2000 km s$^{-1}$ of $z=0.674$, of which 
one (XID 201b) is fainter than R of 24; these 
objects are distributed loosely across a large fraction of the field and 
should thus trace a sheet-like structure. The spike centered on $z=0.734$  
is narrower and includes 16 X-ray sources  within 1000 km s$^{-1}$ 
of the mean redshift, all brighter than R of 24. 
In both structures, about 70\% of the sources are X-ray type-2 AGNs 
(+ X-ray galaxies). The brightest X-ray cluster (XID 594)
 belongs to the $z=0.73$ spike. A few field galaxies, possibly associated 
with this X-ray cluster and other extended X-ray sources (of which XID 645 
at $z = 0.679$), are given at the end of  Table \ref{tblspec}.
The $z=0.67$ and 0.73 structures are also traced by galaxies from the ESO 
K20 survey which covers $\sim$1/10 of the Chandra field 
\citep{cimatti2002a,cimatti2002b}: they are populated by 24 and 47 galaxies
respectively \citep{gilli2003}. The K20 structure at $z=0.73$ is dominated 
by a standard cluster with a central cD galaxy (identified with the 
extended X-ray source XID 566), whereas the K20 galaxies at $z=0.67$ are
uniformaly distributed across the field. There is also evidence of 
higher redshift, narrow spikes in the distribution of the X-ray sources
at $z=1.04$, 1.62 and 2.57; that at $z=1.04$ is also present in the K20  
sample \citep{gilli2003}.

At $z > 2$, there are similar numbers of X-ray type-1 (5) and type-2 (7) 
QSOs. The relative paucity of high $z$ X-ray type-2 AGN (1/6) could arise 
from an observational bias as type-2 sources are optically fainter than 
the type-1 population. 
At $z < 1$, the higher number of X-ray type-2 over type-1 sources is 
mainly due to the large concentration of X-ray type-2 sources within the  
$z=0.67$ and 0.73 structures.

The redshift distribution of the bright sample 
with 60\%  redshift identification completeness   
can be compared to those predicted by models. The X-ray background 
population synthesis models \citep{gilli2001}, based on the AGN/QSO 
X-ray luminosity function and its evolution, 
predict a maximum in the AGN/QSO redshift distribution at $z\sim1.5$. 
Contrary to these expectations, accretion onto black holes is still 
very important at  $z<1$: indeed 88 (68\%) of the 130 
CDFS extragalactic X-ray sources are at $z<1$ and the redshift
distribution peaks around $z\sim0.7$, even if the normal starforming
galaxies are removed from the sample.  
Similar results were found  for the CDFN \citep{barger2002}.  This clearly
demonstrates that the population synthesis models will have to
be modified to incorporate different luminosity functions and
evolutionary scenarios for intermediate-redshift, lower-luminosity AGNs.

Moreover, the CDFS redshift distribution does not confirm the
prediction by \citet{haiman1999}, that a large number ($\sim$100) of 
QSOs at redshifts larger than 5 should be expected in any ultra
deep Chandra survey. The highest redshift in the CDFS thus far
is 3.70, while there are two confirmed and one uncertain high redshift 
sources in the CDFN at $z=4.14, 5.19$ and $z=4.42$,
respectively \citep{barger2002,barger2003a,brandt2001b}, as well as one
QSO at $z=4.45$ in the Lockman Hole \citep{schneider1998}. As our target
selection is based primarily on our R-band imaging, we are suffering
from a bias against z$>$5 objects (the Ly$_\alpha$ emission is 
redshifted out of the FORS R-band at z$\sim$5). Therefore, we may have
a {\em few} QSOs at redshifts larger than 5 in the sample, but we can
be certain that the number of these is on the order of a few. Most of
the X-ray survey area is covered by near-infrared, where objects well
beyond redshift of 15 are detectable. Among the objects covered in the
near-IR, we only find 10 that are detected {\em only} in the near-IR.
Furthermore, 4 of these were still observed spectroscopically,
where we can detect Ly$_\alpha$ emission up to a redshift of
6.5. So in the unlikely case that {\em all} these objects and
five additional objects not detected in optical imaging and
without near-IR coverage are all very high redshift QSOs, we are
still an order of magnitude below the predicted number of such
objects.  This suggests a turn-off of the X-ray selected QSO
space density beyond $z \sim 4$ \citep{hasinger2002,barger2003a}.

\section{OPTICAL AND X-RAY DIAGNOSTICS}

\subsection{\it X-ray and Optical Fluxes\label{optxflux}}

The soft and hard X-ray fluxes versus redshift diagrams are shown in 
Figure \ref{fxz}. The X-ray type-1  and type-2 
populations have similar hard  X-ray fluxes, whereas these two populations 
cover different ranges of soft X-ray fluxes, as can also be seen in 
Figure \ref{fxR}. However, the brighter, rarer objects, $f_{\rm X}$(0.5-2 keV) 
and $f_{\rm X}$(2-10 keV) larger than (1 and 5)$ \times 10^{-14}$
erg cm$^{-2}$ s$^{-1}$ respectively, are dominated by optically 
broad-emission line QSOs (at $z < 2$) as already demonstrated by larger 
samples of luminous X-ray sources detected by ROSAT, Chandra and XMM-Newton 
\citep{lehmann2001,barger2002,mainieri2002}.

\begin{figure}
\plotone{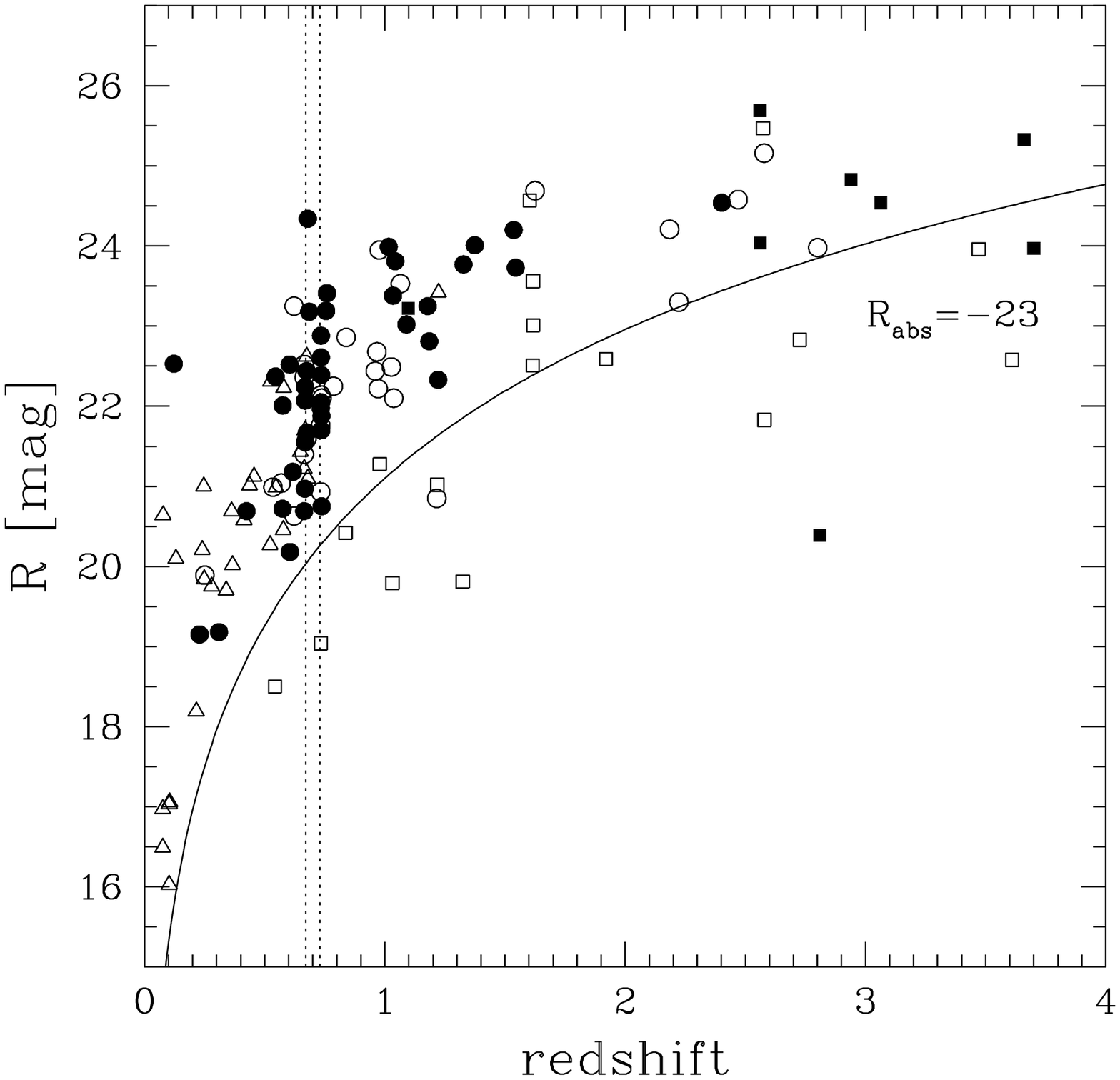}

\caption{R magnitude as a function of redshift for
the CDFS X-ray sources. Symbols are as in Figure \ref{fxR}. 
The dotted lines indicate the two structures around $z=0.67$ and 0.73.
The solid line corresponds to an absolute magnitude of $M_{\rm R}=-23$.
\label{zR}}

\end{figure}

\begin{figure}
\plotone{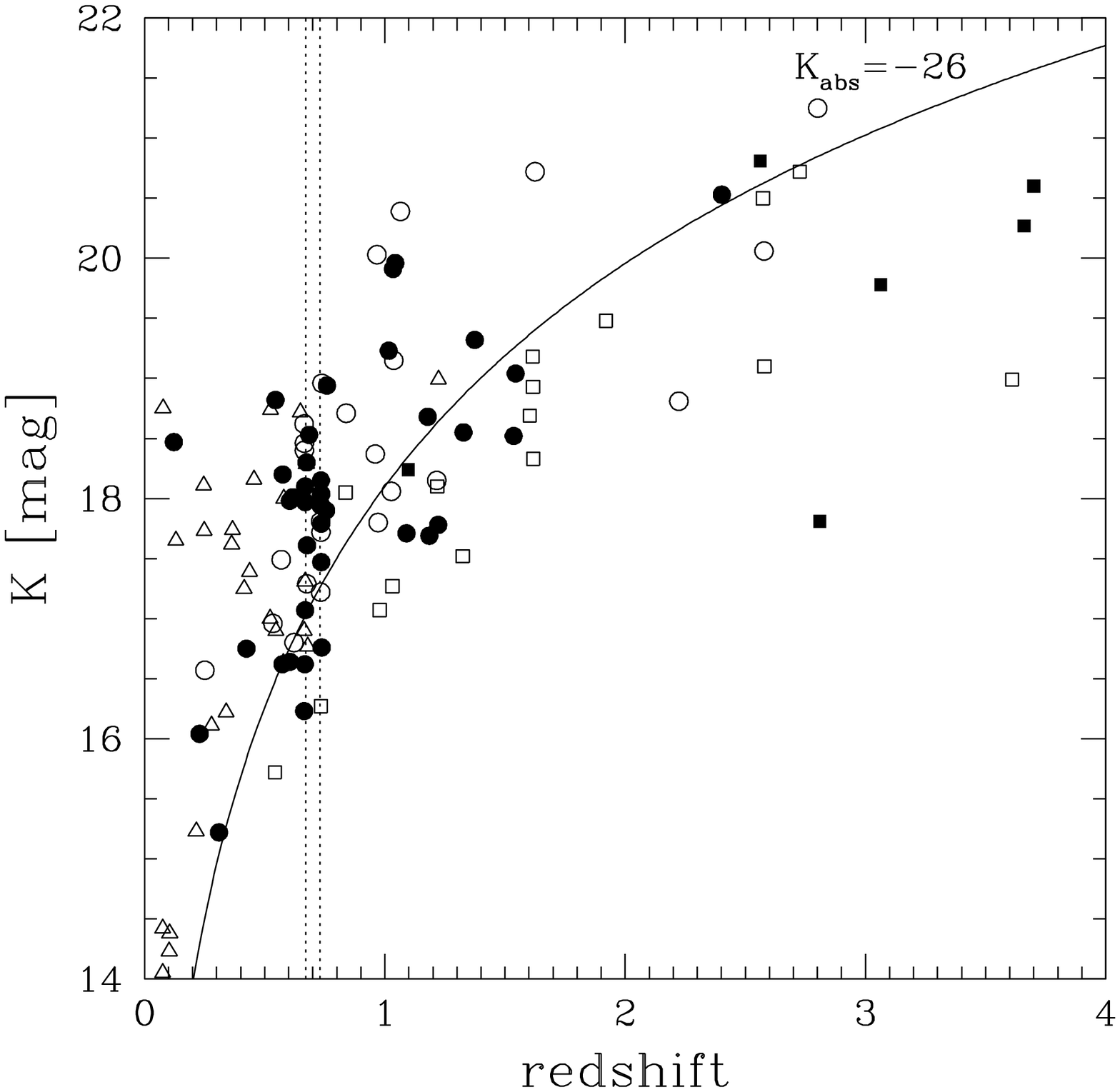}

\caption{K magnitude as a function of redshift. Symbols are as in 
Figure \ref{fxR}.
The solid line corresponds to an absolute magnitude of $M_{\rm K}=-26$.
The brigthest source (in both the K- and R-bands) at $z > 1.5$ is the BAL QSO 
at $z = 2.810$ (XID 62).
\label{zK}}

\end{figure}

The observed R and K magnitudes of the  extragalactic sources versus 
redshift are shown in Figures \ref{zR} and \ref{zK} respectively. 
At $z\gtrsim2$, 
there are seven X-ray type-2 QSOs (XID 27, 54, 57, 62, 112, 202, 263)  
plus one lower X-ray luminosity type-2 AGN (XID 642); except the BAL QSO, 
all have narrow Ly$\alpha$ and C\,{\sc iv} emission, $HR>-0.2$, and faint 
optical magnitudes R $\gtrsim 24.0$. 
There are also six X-ray type-1 QSOs at $z\gtrsim2$ (XID 11, 15, 21, 24, 
68, 117), all but one being otically bright (R $<24$) BLAGN, and
five fainter, lower X-ray luminosity type-1 AGN (XID 87, 89,
230, 563, 901).  The fraction of high-redshift, X-ray 
type-2  QSO+AGN  sources is thus 42\%.
 Moreover, two of the X-ray type-1 QSO/AGN (XID 117, 901), with
narrow Ly$\alpha$ and C\,{\sc iv} emission but $ HR < -0.2$, could be
absorbed X-ray sources since the hardness ratio is not a good
tracer of intrinsic absorption for high redshift sources.  
These results differ from those obtained for the CDFN 2 Msec sample 
\citep{barger2003b} which comprises 26 objects at $z\gtrsim2.0$ (excluding 
the sources with tentative redshifts or complex/multiple structure or 
possible contamination: their Types s and m, respectively) of which 20
are BLAGN, thus an optical type-2 QSO+AGN  fraction of 23\%.
This may arise from an observational selection as 10 (53\%) of the 19  
CDFS sources at $z\gtrsim2$, with secure redshift identification, 
have R $> 24$ as compared to only 2 (8\%) out of 26 CDFN sources.

The segregation between the X-ray type-1 and type-2 QSOs/AGNs seen in 
the R versus $z$ diagram (Figure \ref{zR}) is far less pronounced in
the  K versus $z$ diagram (Figure \ref{zK}).  This is most
likely due to the presence of dust in  X-ray type-2 QSOs/AGNs associated
with the X-ray absorbing material which severely obscures the
 nuclear component, as well as an increased
contribution of the galaxy host light in the K-band relative to
that of the AGN.  The X-ray and optical versus redshift diagrams
(Figure \ref{fxz}, \ref{zR} and \ref{zK})
strongly suggest that the X-ray type-1 and type-2
populations cover roughly the same range of intrinsic
luminosities \citep[see also][]{rosati2002,mainieri2002}.

\subsection{\it X-ray and Optical Colours}

\begin{figure}
\plotone{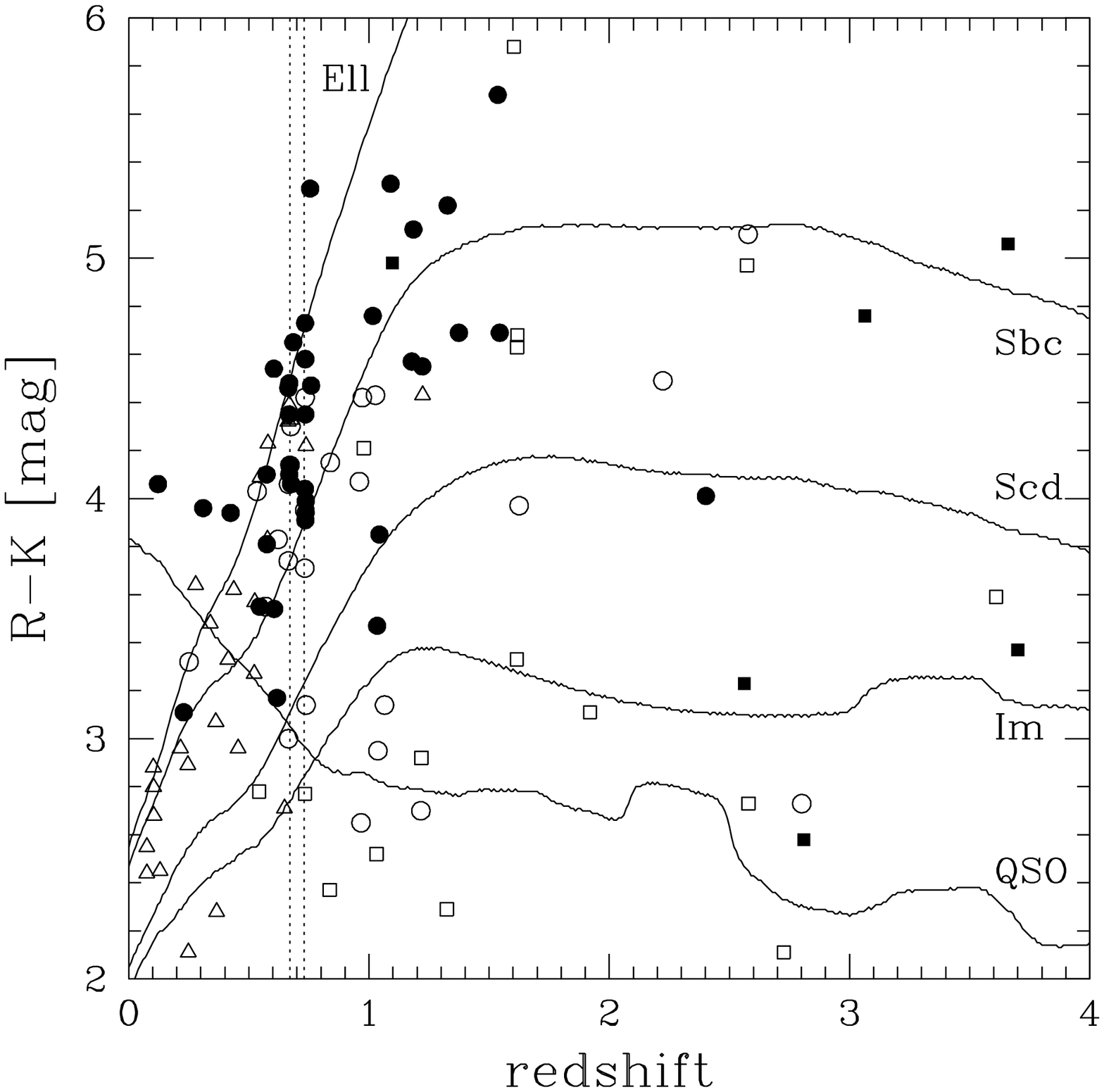}

\caption{R$-$K colour as a function of redshift. Symbols are as in 
Figure \ref{fxR}. 
The five evolutionary tracks correspond to an unreddened QSO, with 
$L = L^{\star}_{\rm B}$, and unreddened elliptical, Sbc, Scd and irregular 
$L^{\star}$ galaxies from the \citet{coleman1980} SED templates.
\label{rkz}}

\end{figure}

A segregation of the X-ray type-1 and type-2 populations is also
present in the R$-$K versus $z$ diagram (see Figure \ref{rkz}),
as first outlined by \citet{lehmann2001} and confirmed by
\citet{mainieri2002}. The deeper Chandra observations reveal
many more X-ray type-2 sources which have optical/near IR colours
dominated by the host galaxy and most of them cluster around the
SED tracks of elliptical and Sbc galaxies at $0.5 < z <1.0$
\citep[see also][]{rosati2002}.  The X-ray type-1 population usually
follows the evolutionary track of an unreddened QSO, except nine  
AGNs/QSOs at $z \gtrsim 1$, all with R$-$K $\gtrsim$ 4. 
This may be due to either an important contribution of the galaxy host 
light in the near IR or obscuration by dust. Among these nine X-ray type-1 
sources, the X-ray luminous, red BLAGN at $z = 1.616$ (XID 67) was 
observed with HST/WFPC2 and is clearly resolved
 with an  elliptical morphology \citep{koekemoer2002}.  A
substantial contribution of the host galaxy could also account
for the red colour of two BLAGN at $z \simeq 1.62$ (XID 46, 101). 
Obscuration by dust associated with the X-ray aborbing material is 
more probable for the remaining six X-ray type-1 sources, of which four 
belong to the HEX optical class and are at $1.6\lesssim z \lesssim 2.6$ 
(XID 31, 117, 563, 901) and two belong to the LEX optical class
and are at $z \simeq 1.0$ (XID 18, 242). 

\begin{figure}
\plotone{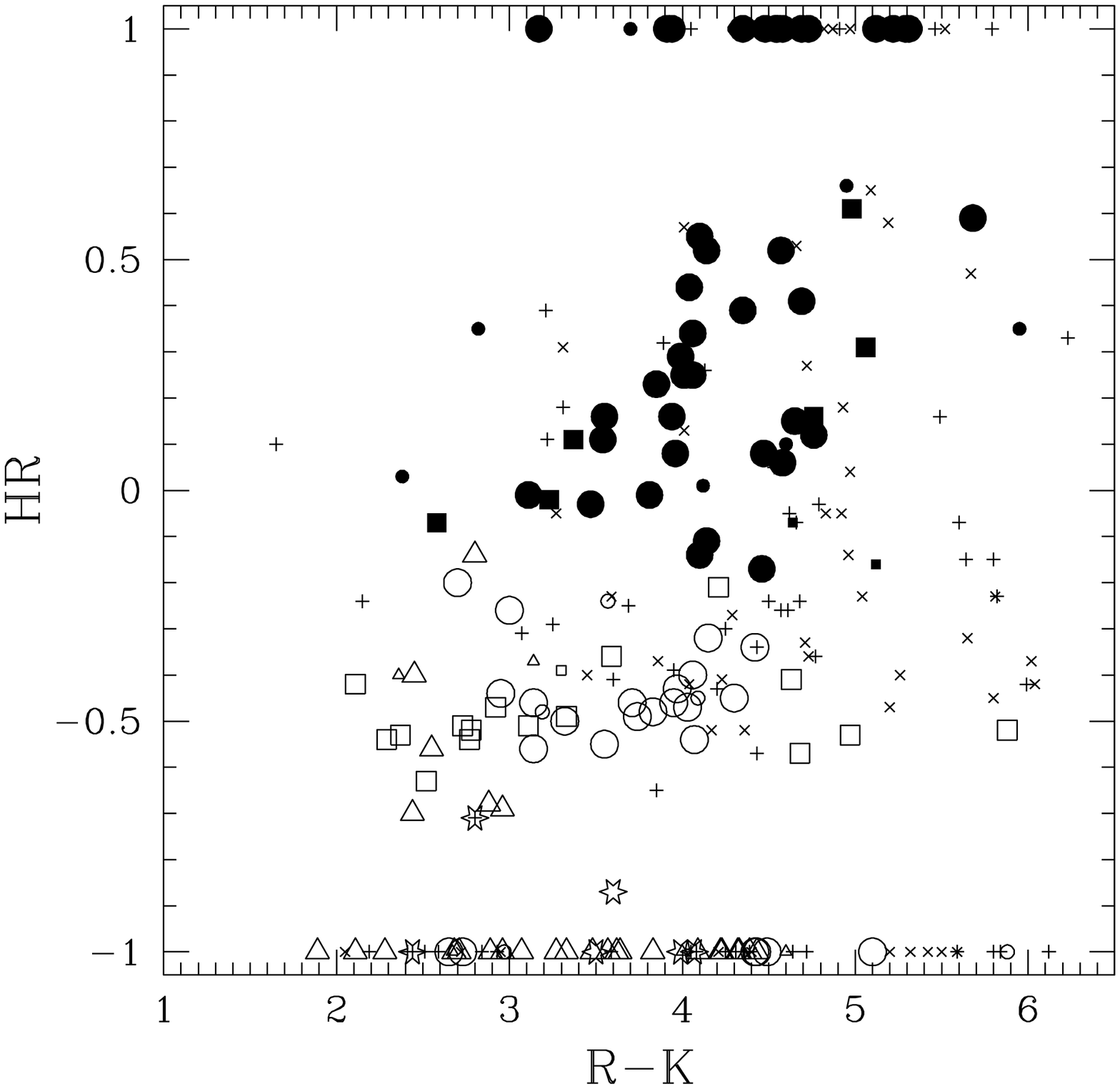}

\caption{The hardness ratio (defined from the net count rates in the 2-10 keV 
and 0.5-2 keV bands: see Section 7) versus R$-$K colour. Symbols are 
as in Figure \ref{fxR} for the reliable $z$ identifications, 
similar symbols but of smaller sizes for the tentative $z$ identifications.
\label{hrrk}}

\end{figure}

For a large fraction of the X-ray sources, there is a relationship between 
the hardness ratio, $HR$, and the  R$-$K colour as shown in Figure \ref{hrrk}. 
The bluer objects are X-ray type-1 QSOs, whereas the redder ones are mostly 
X-ray type-2 AGNs at $z \sim 0.5$ to 1.0. However, the redder objects 
(R$-$K $>$ 4) cover a wide range of $HR$ values, as already noted by 
\citet{franceschini2002}, and they comprise many X-ray type-1 AGNs, including 
the nine objects discussed above, while most of the remaining X-ray type-1 are 
$z < 1$ objects of the optical LEX class.  
To constrain the nature of the redder, X-ray type-1  AGNs requires to 
conduct an 
X-ray spectral analysis (Chandra and XMM-Newton data) of these sources 
(Streblyanskaya, Mainieri et al. in preparation), primarily those with 
secure redshifts, and to derive the morphological properties of their host 
galaxies using the HST observations from the GOODS-ACS Treasury program. 

\subsection{\it X-ray Selected Extremely Red Objects}

The fraction of extremely red objects (EROs: R$-$K $>$ 5.0) among
X-ray sources appears to increase with decreasing optical flux as found 
for a subset of the CDFN X-ray sources \citep{alexander2001} and for 
 Lockman Hole (LH) sources detected by XMM-Newton \citep{mainieri2002}. 
We use the CDFS sample given in Table \ref{tblspec} to confirm this 
trend. There are 151 X-ray sources with bright, R$<$24, counterparts
observed in the R and K bands. Five sources were not detected in 
the K band (K $>$ 20.3), but the upper limits on their R$-$K colours
are smaller than 5. 
The fraction of EROs in this bright optical sample is 10\% (15 objects). 
The fainter optical sample is limited to $24\leq$ R $<26$ to have meaningful 
R$-$K upper limits and it comprises 72 X-ray sources. Most of these faint 
objects do not have spectroscopic redshifts. Thus, in cases of several  
possible counterparts, the brigthest of the best centred counterparts 
was selected. To the 14 EROs with measured R$-$K colours, should be added 
the four objects detected in the K band only (R $>$ 26.3), thus with 
R$-$K $\gtrsim$ 6.0. For the X-ray counterparts not detected in the K band,
all those with $24\leq$ R $<25$ have R$-$K upper limits smaller than 5, 
but among the 10 objects with $25\leq$ R $<26$ only four have 
R$-$K $\lesssim$5.0. The remaining six objects have R$-$K upper limits 
in the range 5.3 to 5.6, but we will consider them as non-EROs in order 
to get a conservative value of the number of EROs among the fainter 
optical sample. The ERO fraction in the R$\geq$24 sample is thus 25\%, or 
2.5 times higher than for the R$<$24 sample. 

Six of the optically bright CDFS EROs have redshift estimates (5 secure), all 
with $z\sim 1$ ($\pm0.3$). One is an X-ray AGN-1 (LEX optical class), and five 
are X-ray AGN-2 (LEX or ABS class) thus with strong optical obscuration 
associated with X-ray absorption. The other nine objects do not show any 
optical emission/absorption feature and all, but one, have $HR<-0.2$.
If they were at $z\sim 1$, they would have luminosities 
$L_{\rm X}$(0.5-10 keV)$>10^{42}$ erg s$^{-1}$. The fraction of hard X-ray 
sources among the optically  bright ERO population is thus 40\%. 
Higher redshift sources are present in the optically  faint ERO sample.
Among the six objects with redshift estimates (4 secure), five are at 
$1.6\lesssim z \lesssim 3.7$ of which three are X-ray luminous QSOs. 
The fraction of hard X-ray, optically  faint EROs is 50\%, but if there
were a majority of high $z$ sources, the bulk of this faint ERO population 
would be heavily absorbed X-ray sources. A similar result was found for 
the LH sources \citep{mainieri2002}. This sample comprises 66 objects 
with measured R$-$K colour (only 20 are fainter than R=24), of which 
18 are EROs. Five EROs (28\%) are not detected in the hard band as compared 
to 27\% and 22\% for the CDFS optically bright and faint ERO samples,  
respectively. The X-ray spectral analysis of the LH sources shows that all, 
but one, of the $HR>-1.0$ sources have high intrinsic absorption:
ten have absorbing column densities $N_{\rm H} > 10^{22}$ cm$^{-2}$ and 
two, without redshift identification, have lower limits (observer frame) 
of $N_{\rm H,min} > 10^{21.5}$ cm$^{-2}$.

Among the X-ray selected EROs, the dominant population appears to be fairly
luminous, absorbed X-ray sources, thus of the X-ray type-2 AGN class, at
intermediate and high redshifts.  This is consistent with the small fraction
(1.5-10\%) of near-IR selected EROs detected in X-ray \citep{cimatti2003}.  
EROs belonging to other classes, elliptical galaxies or dusty starbursts    
\citep{stevens2003}, are also most probably present. Indeed, among the      
optically faint EROs, there are a few sources with optically red and soft   
X-ray spectra (e.g. XID 579b).

Objects of different classes are also found for EROs in the HDFN and 
the Lockman hole \citep{franceschini2002,stevens2003}: 
sources at $z \sim 1$ with SEDs typical of elliptical galaxies, dusty 
starbursts and  $z > 1.5$ absorbed AGN.

\subsection{\it Luminosities}

\begin{figure}
\plotone{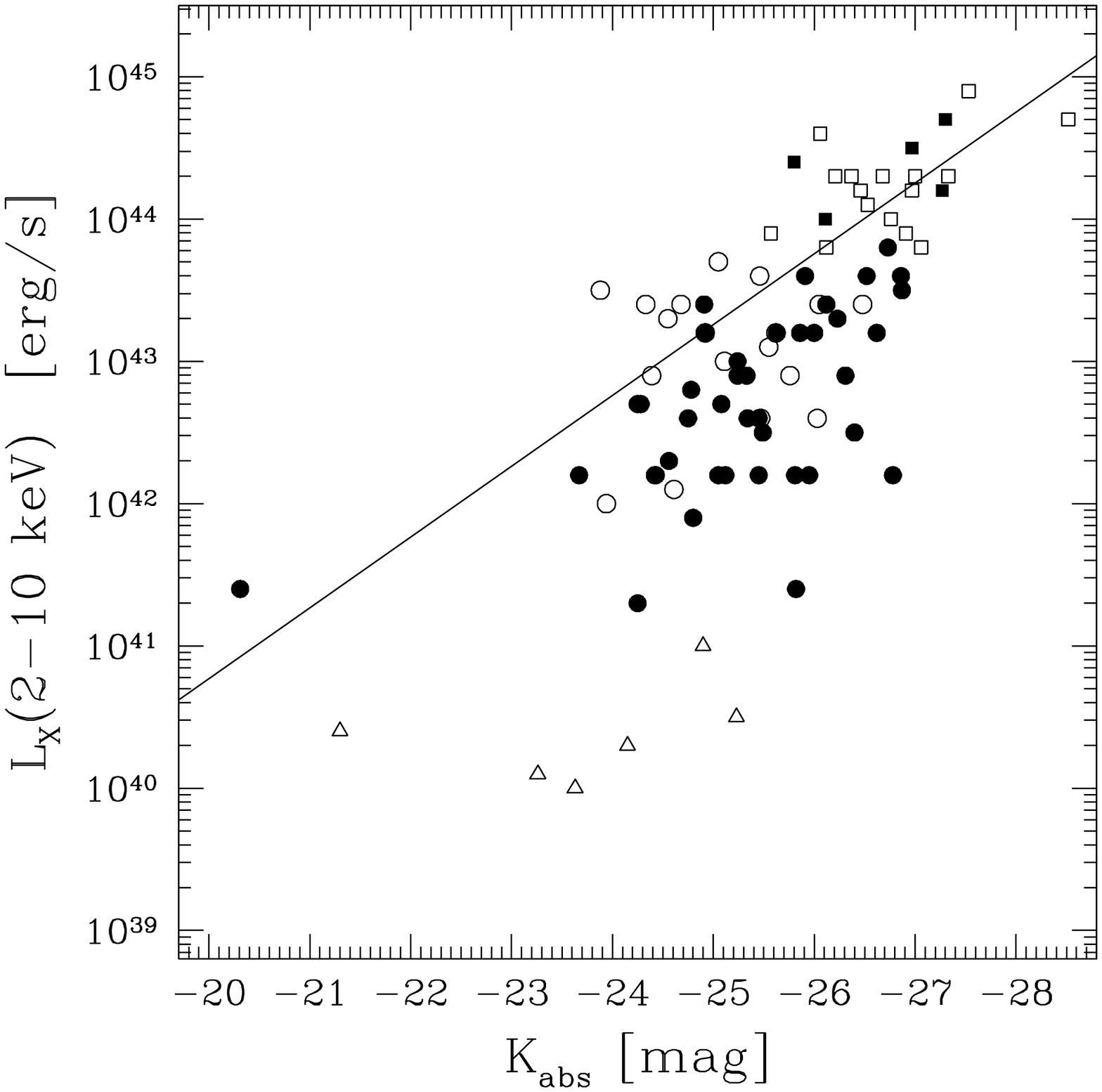}

\caption{The X-ray luminosity in the 2-10 keV band versus the absolute K 
magnitude. Symbols are as in Figure \ref{fxR}. We also convert the
relation found between {\em bulge} luminosity and black hole {\em mass}
\citep{marconi2003}, by assuming that around 40\% of the K-band emission
is coming from the bulge and
an X-ray luminosity of 0.1\% of the Eddington limit (Solid line).
\label{KLx}}

\end{figure}

A trend of increasing hard X-ray luminosity (2-10 keV band) with absolute 
K magnitude can be seen in  Figure \ref{KLx}. 
This trend is not present when the X-ray luminosity in the broad 
0.5-10 keV band is considered instead \citep[see also][]{franceschini2002}.
We also show in this figure the effect expected from the correlation
found between the bulge luminosity and the black hole mass \citep{marconi2003}.
We used very uncertain assumptions to derive this curve. We assumed
that around 40\% of the K-band emission originates in the bulge and we
assumed that the X-ray luminosity is 0.1\% of the Eddington limit luminosity.
The observed X-ray luminosity in the hard band is close to the intrinsic one 
(small K correction for most of the sources) and the reddening correction
for the K absolute magnitude is much smaller in the near-IR than in the 
optical, although it could still be important for the extremely red objects.
We thus expect a tighter correlation between the mid-IR luminosity,  
to be obtained by the Spitzer-GOODS Legacy program,  and the hard X-ray 
luminosity. The trend present in  Figure \ref{KLx} reinforces the
suggestion made above that the X-ray type-1 and type-2 populations cover the 
same range of luminosities, thus trace  similar levels of gravitational
accretion. They differ by either the environment close to the AGN and/or 
the viewing angle to the nucleus, the contribution of the light from 
an early-type  host galaxy, or the dust content and dust-to-gas ratio 
within the host galaxy, or an associated starburst.

The observed hard X-ray luminosity  as a function of redshift is shown in 
Figure \ref{zLx}. There are X-ray luminous X-ray type-1 QSOs down to $z=0.5$,
thus no strong evolution, confirming the findings in the CDFN
\citep{barger2002}.  However, this 
only applies to the X-ray type-1 population, as there is only one (8\%)
X-ray type-2 QSO (XID 51) out of 13 QSOs at $0.5 < z < 2$. 
Two of the X-ray type-1 QSOs show narrow
emission lines only (XID 18, 31) and could be absorbed X-ray sources, as 
indeed confirmed by X-ray spectral analysis of XID 31 ($z=1.603$) which 
is an absorbed source  with $N_{\rm H} = 1.4\times 10^{22}$ cm$^{-2}$ 
(V. Mainieri, private communication). Even including these 
narrow-line QSOs in the type-2 sample would still lead to only 
23\% X-ray type-2 QSOs at lower redshift compared to 54\%  at $z > 2$ 
(see Section \ref{optxflux}). This difference 
(detected at 90\% confidence level) would no longer be as significant 
if sources down to $L_{\rm X}$(2-10 keV) $> 10^{43.5}$ erg s$^{-1}$ 
were considered instead. 
Indeed, the ratio of X-ray type-2/type-1 sources at $0.5 < z < 2$ 
increases with decreasing  hard X-ray luminosity, the type-2 population 
being dominant for $L_{\rm X}$(2-10 keV) $< 10^{43.0}$ erg s$^{-1}$.
This trend is confirmed by the analysis of the 2-10 keV luminosity function 
derived from ASCA, HEAO1 and Chandra surveys  \citep{ueda2003} which 
shows that, at $z < 1$, the percentage of X-ray type-2 AGN 
($N_{\rm H} > 10^{22}$ cm$^{-2}$) decreases with increasing intrinsic 
luminosity from 49\% at $L_{\rm X}$(2-10 keV) $= 10^{43}$ erg s$^{-1}$ 
to 26\% at $L_{\rm X}$(2-10 keV) $= 10^{45}$ erg s$^{-1}$.

The X-ray spectral analysis (absorbing column densities and intrinsic 
X-ray luminosities) of the QSOs+AGNs of the CDFS and CDFN
spectroscopic samples will enable the determination of the cosmic 
evolution of the X-ray type-1 and the X-ray type-2/absorbed sources, thus 
of a possible differential cosmic evolution between these two populations.
 It should be noted that \citet{barger2002}
mention the existence of only two type-2 QSOs at $0.5 < z < 2$ (both at 
$z \approx 1$) while, in their Table 1, there are 12 broad-line QSO+AGN 
sources in the same redshift range, which is consistent with our results.
However, in the CDFN, the fraction of type-2 QSOs is small at both 
intermediate and high (see Section \ref{optxflux}) redshifts.

\section{SUMMARY AND OUTLOOK }

We presented a catalog of 137 secure and 24 tentative spectroscopic
identifications of the 349 X-ray objects (including 3 new, faint
sources) in the CDFS field, based on our survey using the VLT. Our 
spectroscopic survey is 40\% complete considering the whole X-ray catalog, 
and 70\% complete if we consider the subset in the central 8\arcmin~ radius 
with optical counterparts at R$<$24. This can compared to 
the somewhat higher spectroscopic completeness achieved in the Chandra Deep 
Field North identification programme \citep{barger2002}, where the 
corresponding fractions are 49\% and 78\%, respectively. 
Very recently, optical identifications have also been presented
for the 2 Msec observation of the HDFN \citep{barger2003b}, which
reach a completeness as high as 87\% at R$<$24. At fainter optical 
magnitudes (R$>$24), however, the fraction of reliable spectroscopic 
identifications is larger for the CDFS compared to the HDFN. This is 
becoming important in particular, when comparing the fraction of 
X-ray type-2 QSOs at these faint magnitudes (see below). 
   
We proposed a new, objective and simple scheme, based on
X-ray luminosity and hardness ratio, to classify objects into 
X-ray type-1 (unabsorbed) and X-ray type-2 (absorbed) AGN. Hard ($HR>-0.2$)
sources are classified as X-ray type-2 AGN or QSO, depending on their X-ray 
luminosity. Soft sources ($HR\leq-0.2$) are classified as X-ray galaxies, 
X-ray type-1 AGN or QSO, depending on their X-ray luminosity. At high optical 
and X-ray luminosities, this classification scheme is largely coincident 
with the classical AGN classification purely based on optical spectroscopic 
diagnostics. However, as soon as the integrated light of the host 
galaxy becomes larger than the optical emission of the AGN 
nucleus, the optical classification breaks down. Consequently,
we are classifying many more objects as AGN, than would be 
selected in optical samples. An additional advantage of our proposed
classification scheme is that it only relies on X-ray fluxes and redshift
(to calculate $L_X$). So far we only used optical spectroscopy to derive
the redshift, but our scheme can use photometric redshift techniques, thus
going significantly beyond the capabilities of optical spectroscopy. Indeed,
using photo-$z$ techniques, more than 95\% of the CDFS sources can be
identified in our scheme (Mainieri, private communication).

\begin{figure}
\plotone{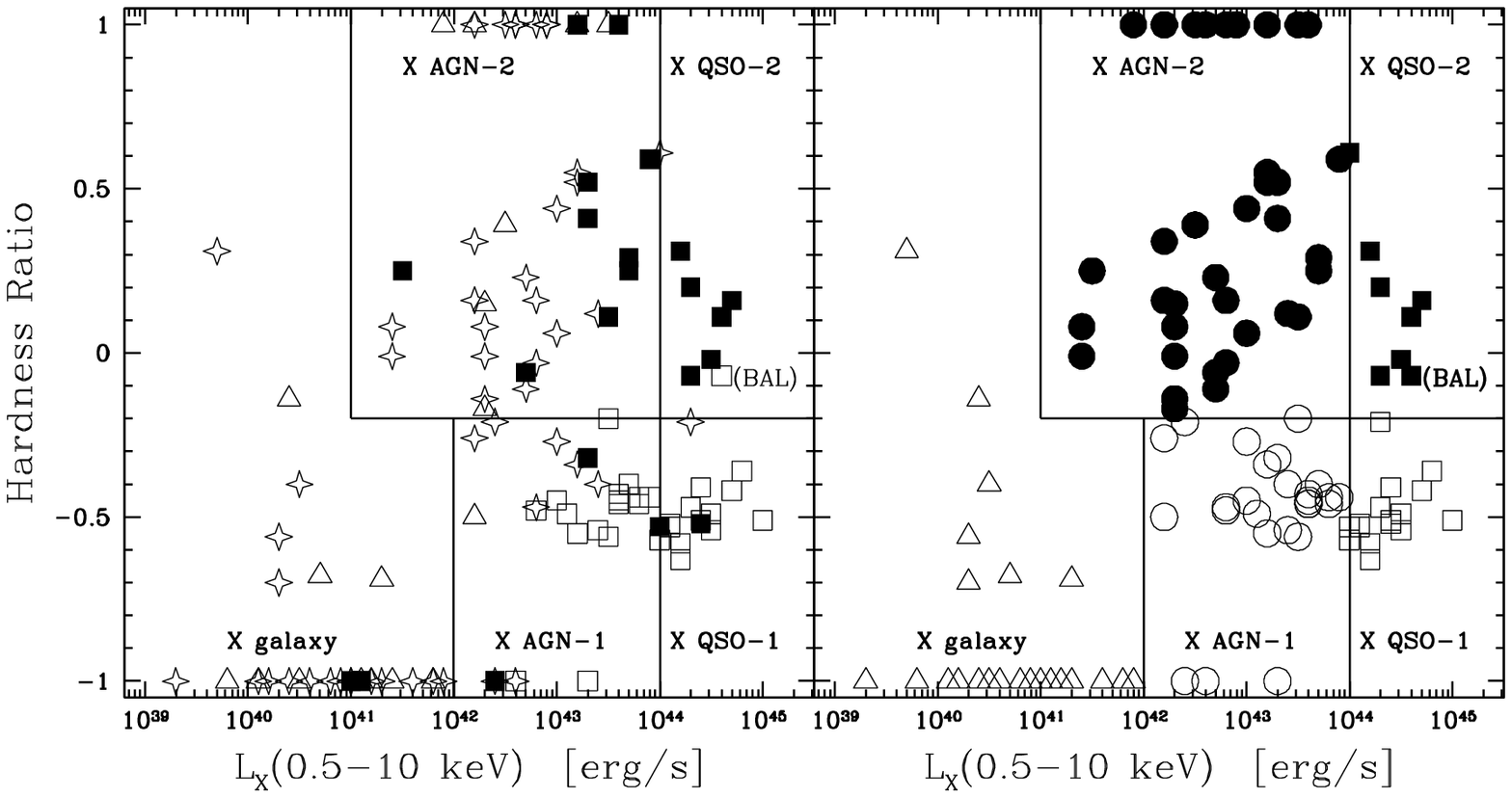}

\caption{Hardness ratio versus observed X-ray luminosity in the
0.5-10 keV band. Symbols in the left panel show the classical optical
classification (empty squares: BLAGN, filled squares: HEX, diamonds: LEX,
triangles: ABS). Symbols in the right panel are as in see Figure \ref{fxR}
and follow  the X-ray classification presented in Section 
\ref{zestimate}\label{hrlx}. The BAL QSO (XID 62) is also marked.}

\end{figure}

\begin{figure}
\plotone{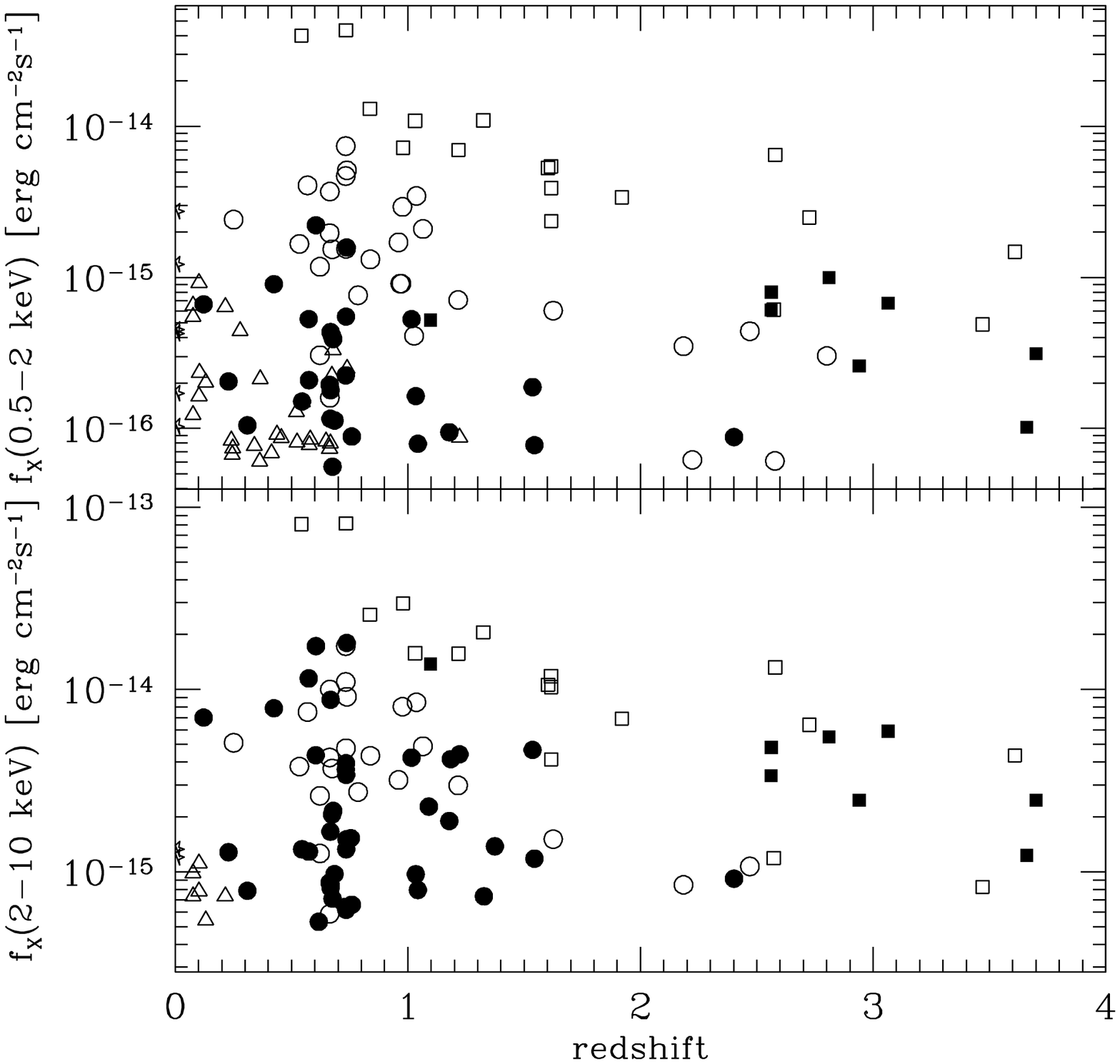}

\caption{X-ray flux of the CDFS sources as a function of redshift for
the 0.5-2 keV band (top) and 2-10  keV band (bottom). 
Symbols are as in Figure \ref{fxR}. 
\label{fxz}}

\end{figure}

\begin{figure}
\plotone{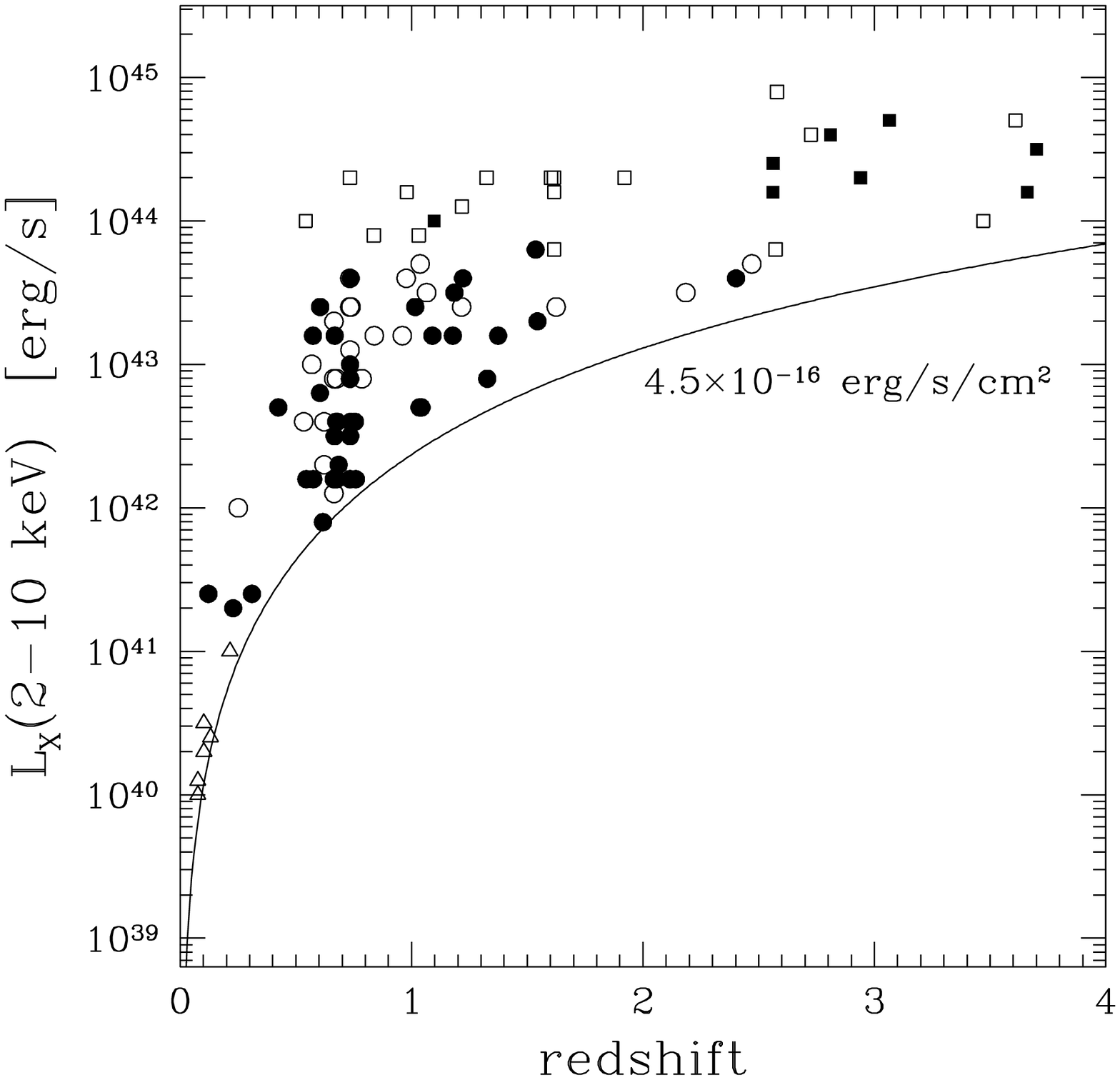}

\caption{The X-ray luminosity in the 2-10 keV band versus 
redshift. Symbols are as in Figure \ref{fxR}. 
\label{zLx}}

\end{figure}

We have spectroscopically identified a sample of 8 secure and 2
tentative high-luminosity X-ray sources with significant absorption, 
our X-ray type-2 QSO class. Nine ($^{+4.1}_{-3.0}$: 1$\sigma$ errors) 
of these sources are in the redshift range $2<z<4$ and their optical
spectra are dominated by strong, narrow high excitation UV permitted lines,
very similar to the prototypical object CDFS-202 \citep{norman2002}.
In contrast, the spectroscopic sample existing in the HDFN so far
\citep{barger2002} only contains 2 ($^{+2.6}_{-1.3}$) similar objects 
(HDFN $\# 184$ and $\# 287$). 
This difference may be due to the fact that our spectroscopy
is pushing about one magnitude deeper than the HDFN spectroscopy in a part 
of the field. However, we can not exclude true cosmic field-to-field 
variations in the number of X-ray type-2 QSOs. The fraction of X-ray type-2 
to the total AGN population shows a significant variation with observed X-ray 
luminosity, consistent with, but even somewhat stronger than the trend 
found from ASCA surveys in the 2-10 keV band \citep{ueda2003}: the X-ray 
type-2 fraction decreases from $75\pm8\%$ (8 type-1 vs. 24 type-2 AGN) 
in the luminosity range $10^{42-43}$~erg~s$^{-1}$, over $44\pm8\%$ 
(20 vs. 16 AGN) at luminosities $10^{43-44}$~erg~s$^{-1}$, to $33\pm10\%$ 
(16 vs. 8) at $10^{44-45}$~erg~s$^{-1}$
(see also Figure \ref{hrlx}). This behaviour can probably explain some 
of the evolutionary trends apparent in Figures \ref{fxz} and \ref{zLx}.  

We found spectroscopic 
evidence for two large-scale structures in the field, predominantly populated
by X-ray type-2 AGN but also X-ray type-1 AGN and normal galaxies: one at
$z = 0.734$ has a fairly narrow redshift distribution and comprises two 
clusters/groups of galaxies centered on extended X-ray sources. The redshift
distribution of the second one at $z = 0.674$ is broader (velocity space) 
and traces a sheet-like structure. A detailed comparison with the redshift 
spikes in a NIR-selected (K20) sample of galaxies in the same field has been 
performed by \citep{gilli2003}. Similar, but much less pronounced
redshift spikes have also been observed in the HDFN at redshifts around
$z =  0.843$ and $z = 1.018$ by \citep{barger2002}.  
AGN therefore trace large-scale structures as do normal galaxies.
Further studies on larger samples are required to investigate, whether
AGN are more strongly clustered than normal galaxies \citep{gilli2003} 
and, whether X-ray type-2 AGN are indeed clustering stronger than 
X-ray type-1 AGN, as indicated by the CDFS results.

However, the objects in these spikes do not dominate the sample.
The observed AGN redshift distribution peaks at $z\sim0.7$, even if 
the objects in the spikes and also the normal, starforming galaxies
are removed.  Compared to the pre-Chandra and XMM-Newton predictions
of population synthesis models of the X-ray background \citep{gilli2003},
there is an excess of $z<1$ AGN, even taking into account the 
spectroscopic incompleteness of the sample. These models will 
therefore have to be modified to incorporate different luminosity functions and
evolutionary scenarios for intermediate-redshift, lower-luminosity AGNs.
  
It will be interesting to study the correlation of active galaxies to 
field galaxies in the sheets and investigate the role that
galaxy mergers play in the triggering of the AGN activity.
Finally, there may be a relation between the surprisingly low
redshift of the bulk of the Chandra sources, the existence of
the sheets at the same redshift and the strongly evolving
population of dusty starburst galaxies inferred from the ISO
mid-infrared surveys \citep{franceschini2002}.

The Chandra Deep Field South has been
selected as one of the deep fields in the Spitzer legacy programme
Great Observatories Origins Deep Survey (GOODS).  GOODS will
produce the deepest observations with the Spitzer IRAC instrument
at 3.6-8$\mu$m  and with the MIPS instrument at 24$\mu$m  
over a significant fraction of the CDFS  \citep[see][]{fosbury2001}. 
The same area has already been covered by
an extensive set of pointings with the new Advanced Camera for
Surveys (ACS) 
of the Hubble Space Telescope in BVIz to near HDF depth. Following 
up the deep
EIS survey in the CDFS, ESO has undertaken a large program to image
the GOODS area with the VLT to obtain deep JHKs images in some
32 ISAAC fields. 
A small spot inside the CDFS has also been selected as the location of 
the HST ACS ultradeep field (UDF), aiming at roughly two 
magnitudes fainter than the Hubble Deep Fields, over a substantially 
larger area. 
An even larger field than the CDFS has been surveyed with the 
HST ACS program GEMS and has also been covered by multiband 
optical photometry as part of the COMBO-17 survey \citep{wolf2003}.  
The next step in the optical identification and classification 
work is to use the extremely deep HST ACS and VLT ISAAC (or EIS SOFI) data 
provided by GOODS and the narrow band photometry provided by 
COMBO-17 to obtain multicolour photometric redshifts for the 
objects not covered by and/or too faint for our spectroscopic 
identification programme (Zheng et al., in preparation).
 
Additional X-ray information in an area wider than the CDFS is 
existing from a deep XMM-Newton pointing of $\sim$400 ksec  exposure time
(PI: Bergeron).  The already existing Chandra 
Megasecond coverage will be widened and deepened with four 
additional 250 ksec ACIS-I pointings (PI: Brandt). 
The multiwavelength coverage of the field is complemented
by deep 20 cm radio data from the VLA and ATCA.
The CDFS will therefore ultimately be one
of the patches in the sky  providing a combination of the widest
and deepest coverage at all wavelengths and thus a legacy for
the future.



\acknowledgments

This publication makes use of data products from the Two Micron
All Sky Survey, which is a joint project of the University of
Massachusetts and the Infrared Processing and Analysis
Center/California Institute of Technology, funded by the
National Aeronautics and Space Administration and the National
Science Foundation. Three of our redshifts have been obtained from the 
2dFGRS public dataset.





\clearpage


\clearpage

\clearpage

\clearpage

\clearpage

\begin{figure}
\caption{For these figures, see http://www.mpe.mpg.de/CDFS.
Finding charts and VLT spectra of CDFS sources. Images
are based on our FORS R-band imaging. The circles indicate the
positional error \citep{giacconi2002}. The images are $20''\times20''$ in
size. If multiple optical counterparts were considered, they are marked
on the finding charts. Next to the finding charts we show {\em all}
spectroscopic observations available for the object. On the horizontal
axis both the observed (bottom) and rest frame (top -- if available)
wavelength is shown in {\AA} units. The vertical axis is the measured
flux, $f_\lambda$ in $10^{-18}$ erg cm$^{-2}$ s$^{-1}$ \AA$^{-1}$ units. 
Important emission (above the spectra) and absorption (below the spectra) 
features used for identification are also marked.
\label{spectra}}
\end{figure}

\clearpage

\begin{figure}
\caption{For this figure, see http://www.mpe.mpg.de/CDFS/.
Extended object 138\label{cdf138}. The image covers $60\times60$ arcsec.}
\end{figure}

\clearpage 

\begin{figure}
\caption{For this figure, see http://www.mpe.mpg.de/CDFS/.
Extended object 249\label{cdf249}. The image covers $60\times60$ arcsec.}
\end{figure}

\clearpage

\begin{figure}
\caption{For this figure, see http://www.mpe.mpg.de/CDFS/.
Extended object 560\label{cdf560}. The image covers $60\times60$ arcsec.}
\end{figure}

\clearpage

\begin{figure}
\caption{For this figure, see http://www.mpe.mpg.de/CDFS/.
Extended object 566\label{cdf566}. The image covers $60\times60$ arcsec.}
\end{figure}

\clearpage

\begin{figure}
\caption{For this figure, see http://www.mpe.mpg.de/CDFS/.
Extended object 594\label{cdf594}. The image covers $60\times60$ arcsec.}
\end{figure}

\clearpage

\begin{figure}
\caption{For this figure, see http://www.mpe.mpg.de/CDFS/.
Extended object 645\label{cdf645}. The image covers $60\times60$ arcsec.}
\end{figure}

\clearpage 

\clearpage 

\clearpage 

\clearpage 

\clearpage 

\clearpage 

\clearpage 

\clearpage 

\clearpage 

\clearpage 

\clearpage 

\clearpage 

\begin{figure}

\caption{For these figures, see http://www.mpe.mpg.de/CDFS/.
Finding charts and VLT spectra of the field sources. Images
are based on our FORS R-band imaging. The images are $20''\times20''$ in
size.  Next to the finding charts we show {\em all}
spectroscopic observations available for the object. On the horizontal
axis both the observed (bottom) and rest frame (top -- if available)
wavelength is shown in {\AA} units. The vertical axis is the measured
flux, $f_\lambda$ in $10^{-18}$ erg cm$^{-2}$ s$^{-1}$ \AA$^{-1}$ units. 
Important emission (above the spectra) and absorption (below the spectra) 
features used for identification are also marked.
\label{fieldPlots} 
}

\end{figure}

\clearpage


\begin{deluxetable}{r@{~~~~}r@{}r@{~~~}r@{~~~}r@{~~}r@{~~~}r@{~}r@{~~~~}r@{~~}r@{~~~~}r@{~}r@{~~~}}
\tabletypesize{\tiny}
\tablewidth{0pt}
\tablecaption{X-ray sources not yet published
in \cite{giacconi2002}.\label{newsrc}}
\tablewidth{0pt}
\tablehead{
\colhead{XID\tablenotemark{a}} &
\colhead{CXO CDFS\tablenotemark{b}} &
\colhead{RA (J2000)\tablenotemark{c}} &
\colhead{Dec (J2000)\tablenotemark{d}} &
\colhead{Soft Cts.\tablenotemark{e}} &
\colhead{Hard Cts.\tablenotemark{f}} &
\colhead{Exp Time\tablenotemark{g}} &
\colhead{Soft Flux\tablenotemark{h}} &
\colhead{Error\tablenotemark{i}} &
\colhead{Hard Flux\tablenotemark{j}} &
\colhead{Error\tablenotemark{k}} &
\colhead{HR\tablenotemark{l}}
}
\startdata
 901 & J033235.8-274917 & 03 32 35.78 & -27 49 16.82 & 11.0$\pm$5.0  & $<$9.0       &  828.4(857.7)  &     6.1e-17 &  2.9e-17 & $<$3.1e-16 &         & -1 \\
 902 & J033222.1-275113 & 03 32 22.08 & -27 51 13.05 &       $<$7.0  & 12.1$\pm$6.2 &  792.3(804.1)  &  $<$3.9e-17 &       -- &    4.5e-16 & 2.0E-16 & +1 \\
 903 & J033226.0-274049 & 03 32 25.97 & -27 40 49.21 & 22.5$\pm$10.4 & $<$3         &  824.3(820.2)  &     1.3e-16 & 0.6e-16  &   $<1$e-16 &         & -1 \\
\enddata
\tablenotetext{a}{Unique Detection ID}
\tablenotetext{b}{IAU Registered Name, based on X--ray
coordinates}
\tablenotetext{c}{Right Ascension}
\tablenotetext{d}{Declination}
\tablenotetext{e}{Net Counts in soft (0.5 - 2 keV) band}
\tablenotetext{f}{Net Counts in hard (2 - 10 keV) band}
\tablenotetext{g}{Effective exposure time in the soft(hard) band, in
kiloseconds}
\tablenotetext{h}{Flux in soft (0.5 - 2 keV) band, c.g.s. units}
\tablenotetext{i}{Error on soft flux}
\tablenotetext{j}{Flux in hard (2 - 10 keV) band, c.g.s. units}
\tablenotetext{k}{Error on hard flux}
\tablenotetext{l}{Hardness Ratio, defined as (H - S)/(H + S) where H and S
are the net counts in the hard and soft bands respectively}
\end{deluxetable}

\clearpage

\begin{deluxetable}{lcccc}
\tabletypesize{\scriptsize}
\tablecaption{X-ray-to-optical flux ratio ranges for different classes of 
objects \label{fxfo_classes}}
\tablewidth{0pt}
\tablehead{
	\colhead{Object class} & 
        \colhead{$\log(f_{0.3\dots3.5}/f_V)$} & 
	\colhead{$f_X$-shift} & 
	\colhead{V-R} &
        \colhead{$\log(f_{0.5\dots2.0}/f_R)$}
}
\startdata
AGN		& $-$1.0\dots$+$1.2 & 0.25\dots 0.33& 0.0\dots 1.0 & $-$1.4\dots$+$1.1\\
BL Lac		& $+$0.3\dots$+$1.7 & 0.25\dots 0.33& 0.0\dots 1.0 & $-$0.1\dots$+$1.6\\
Clusters	& $-$0.5\dots$+$1.5 & 0.18\dots 0.25& 0.1\dots 0.4 & $-$0.6\dots$+$1.6\\
Galaxies	& $-$1.8\dots$-$0.2 & 0.10\dots 0.30& 0.1\dots 1.0 & $-$2.2\dots$+$0.0\\
M stars		& $-$3.1\dots$-$0.5 & 0.05\dots 0.20& 0.6\dots 1.0 & $-$3.4\dots$-$0.4\\
K stars		& $-$4.0\dots$-$1.5 & 0.05\dots 0.20& 0.4\dots 0.6 & $-$4.1\dots$-$1.3\\
G stars		& $-$4.3\dots$-$2.4 & 0.05\dots 0.20& 0.3\dots 0.5 & $-$4.4\dots$-$2.2\\
B-F stars	& $-$4.6\dots$-$3.0 & 0.05\dots 0.20&-0.5\dots 0.3 & $-$4.6\dots$-$2.5\\
\enddata


\tablecomments{The typical X-ray-to-optical flux ratios in our
new bands (0.5\dots2 keV in X-ray, R-band in optical) as derived
from the previously determined values \citep{stocke1991}. The
$f_X$-shift is the typical shift of X-ray flux (in $\log_{10}$
units) due to the narrower energy range, assuming typical X-ray
spectra for the class of objects. V-R is the typical optical
color range in the class. The last column includes our new
normalization of the X-ray-to-optical flux ratio, as discussed
in the text (Section \ref{optid}).  In the case of clusters, the
magnitude refers to the brigtest cluster galaxy (BCG).}

\end{deluxetable}

\clearpage

\begin{deluxetable}{lrrrrrrc}
\tabletypesize{\scriptsize}
\tablecaption{The location of lines in second order diffraction
\label{2ndorder}}
\tablewidth{0pt}
\tablehead{
	\colhead{Line}         & 
        \colhead{Position$^1$} & \colhead{Flux$^1$} & \colhead{FWHM$^1$} &
        \colhead{Position$^2$} & \colhead{Flux$^2$} & \colhead{FWHM$^2$} &
        \colhead{Filter}    \\
	\colhead{(\AA)}        &
        \colhead{(\AA)}        & \colhead{(ADU)}    & \colhead{(\AA)}    &
        \colhead{(\AA)}        & \colhead{(ADU)}    & \colhead{(\AA)}    &
	\colhead{(Bessel)} 
}
\startdata
 3888.6 & 3887 &    13847 & 26 &   7462 &   3632 &  30 & U \\
 3650.1 & 3648 &    90265 & 30 &   6954 &  36364 &  35 & U \\
 3650.1 & 3649 &    17119 & 27 &   6957 &   8380 &  34 & B \\
 3888.6 & 3888 &    79904 & 27 &   7465 &  19142 &  30 & B \\
 4046.6 & 4047 &   346283 & 30 &   7801 &  44653 &  30 & B \\
 4358.3 & 4359 &   795630 & 28 &   8457 &  51040 &  30 & B \\
 4471.5 & 4472 &    97692 & 27 &   8694 &   5151 &  30 & B \\
 5015.7 & 5014 &   172340 & 28 &   9878 &   1946 &  33 & V \\
\enddata


\tablecomments{The location, apparent width and apparent strength of
selected arc lines in first and second order diffraction in FORS, using
the 150I grism. Flux values are in instrumental counts (ADU). The {\em
Filter} column indicates which (Bessel) filters we used for the particular
line. The observed location of the first order diffraction was used to
test that the filters do not introduce significant shifts in our
wavelength solution.}

\end{deluxetable}

\clearpage

\begin{deluxetable}{clcccl}
\tabletypesize{\scriptsize}
\tablecaption{Summary of spectroscopic observations.\label{tblobs}}
\tablewidth{0pt}
\tablehead{
	\colhead{Mask}       & \colhead{Exposure}   & \colhead{Slit}      &
	\colhead{seeing}     & \colhead{Slitloss}   & \colhead{Night}       \\
        \colhead{}           & \colhead{(s)}        & \colhead{(arcsec)} &
	\colhead {(arcsec)} & \colhead{\%}         & \colhead{}
}
\startdata
6+7         & 3$\times$1800, 3$\times$1800            & 1.2 & 0.8/0.6 & 40  & 2000 Oct 27-28\\
15          & 4$\times$1800+1300                      & 1.2 & 0.7/0.8 & 45  & 2000 Oct 27-28\\
22+23+24    & 3$\times$1800, 1800, 2$\times$1800      & 1.2 & 1.0/0.9 & 50  & 2000 Oct 28-29\\
28+29       & 2$\times$1800, 3$\times$1800            & 1.2 & 0.6/0.9 & 40  & 2000 Oct 28-29\\
36+39+40    & 1800, 1800, 4$\times$1800               & 1.2 & 1.3/0.6 & 65  & 2000 Oct 29-30\\
46+47       & 4$\times$1800, 1800+900                 & 1.2 & 0.5/0.7 & 35  & 2000 Oct 29-30\\
78          & 6$\times$1800+945                       & 1.2 & 0.5/0.6 & N/A & 2000 Nov 24-25\\
82          & 5$\times$1800                           & 1.2 & 0.6/0.6 & 35  & 2000 Nov 23-24\\
84          & 6$\times$1800                           & 1.2 & 0.9/0.8 & 45  & 2000 Nov 23-24\\
86          & 3$\times$1800+1535                      & 1.2 & 1.1/1.0 & 40  & 2000 Nov 24-25\\
88          & 1200                                    & 1.2 & 0.8/0.6 & 80  & 2000 Nov 23-24\\
89          & 1200                                    & 1.2 & 0.8/0.6 & 75  & 2000 Nov 23-24\\
90          & 1200                                    & 1.2 & 1.3/1.1 & 90  & 2000 Nov 23-24\\
99          & 1800                                    & 1.2 & 0.9/1.3 & 50  & 2000 Nov 24-25\\
119+120+121 & 2700, 2$\times$2700, 2$\times$2700+3600 & 1.0 & 0.5/0.7 & 70  & 2001 Sep 20-21\\
122         & 4$\times$2700                           & 1.0 & 0.6/0.8 & 70  & 2001 Sep 18-19\\
134         & 2500+2700                               & 1.4 & 0.9/0.8 & 70  & 2001 Sep 17-18, 19-20\\
137         & 1800+2700                               & 1.4 & 1.0/1.0 & 60  & 2001 Sep 18-19, 19-20\\
138+139     & 3$\times1800$, 2$\times$2700            & 1.0 & 0.5/0.6 & 55  & 2001 Sep 17-18\\
146         & 3$\times$2700                           & 1.4 & 0.7/0.9 & 60  & 2001 Sep 19-20\\
MXU2.1      & 9$\times$1800+2400                      & 1.0 & 0.7/0.9 & N/A & 2001 Nov 13-14, 14-15\\
MXU4.1      & 12$\times$1800+1134                     & 1.0 & 0.9/0.8 & N/A & 2001 Nov 12-13\\
MXU5.1      & 4$\times$1800                           & 1.0 & 0.5/0.7 & 70  & 2001 Nov 14-15\\
MXU11.1     & 9$\times$1800+2400                      & 1.0 & 0.6/0.8 & sl  & 2001 Nov 13-14\\
MOS11.1     & 4$\times$1800                           & ?.? & 1.2/1.0 & 75? & 2001 Nov 11-12\\
MXU1.1      & 2$\times$2100                           & ?.? & 0.5/0.7 & sl  & 2001 Dec 18-19\\
\enddata


\tablecomments{Seeing values refer to seeing directly measured on
the acquisition image (first value) and the seeing measured by
the DIMM seeing monitor during the observations (second number).
The slitloss is estimated from relatively bright point sources
(if observed) by comparing synthetic magnitudes calculated 
from the flux calibrated spectra and broad-band magnitudes.
This value is only to be used to asses the overall quality of 
each mask as it does not take into account the extend and
centering of individual objects (see Section \ref{abscal}).
}

\end{deluxetable}





\clearpage

. Make sure there is at least one \tablenotemark

\tablecomments{
ref 1: A. Cimatti \& R. Gilli, private communication,
ref 2: paper I: Giacconi et al. 2001, ApJ 551,624,
ref 3: C. Wolf, private communication,
ref 4: \citet{daddi2003}\\
See Section \ref{zestimate} for explanation of columns.\\
\ *: extended X-ray source.\\
See Table \ref{lineratios} for diagnostics on line ratios.
}

\end{deluxetable}





\clearpage

\begin{deluxetable}{lccccc}
\tabletypesize{\scriptsize}
\tablecaption{Line ratios. \label{lineratios}}
\tablewidth{0pt}
\tablehead{
\colhead{No} & 
\colhead{$\log{\left[{\rm O III}\right]\over {\rm H\beta}}$} & 
\colhead{$\log{\left[{\rm S II}\right]\over {\rm H\alpha}}$} & 
\colhead{$\log{\left[{\rm N II}\right]\over {\rm H\alpha}}$} & 
\colhead{$\log{\left[{\rm O I}\right]\over {\rm H\alpha}}$} &
\colhead{Class}\\
\colhead{} & 
\colhead{[5007\AA/4861\AA]} & 
\colhead{[6716+6731\AA/6563\AA]} &
\colhead{[6583\AA/6563\AA]} &
\colhead{[6300\AA/6563\AA]} &
\colhead{}
}
\startdata
 41 &  0.98 &  -    &    -  &   -  & Seyfert II\\
 75 &  0.56 &  -    &    -  &   -  & - \\
155 &  0.94 &  -    &    -  &   -  & Seyfert II\\
175a &  0.25 & $-$0.58 & $-$0.60 &   -  & NELG \\
190 &  1.00 &  -    &    -  &   -  & Seyfert II\\
201b &  0.88 &  -    &    -  &   -  & Seyfert II\\
521 &  0.26 &  -    &   $^x$   & $-$0.98 & LINER \\
525 &$>$1.0 &  -    &   \ \ \ \ 0.04$^x$  &   - & Seyfert II\\
538 &  0.73 &  -    & $-$0.03 &   -  & Seyfert II\\
553a &  0.56 & $-$0.94 &    -  &   -  & HII region\\
565 &  0.07 & $-$0.51 & $-$0.46 &   - &   NELG\\
567a & $-$0.11 &  -    &    -  &   -  & - \\
582b & $<$0.0  &  -    & $-$0.39 &   -  & - \\
627 &  0.28 & $-$0.61 &    -  &   -  &  NELG\\
652 &   -   & $-$0.48 &    -  &   -  & - \\
\enddata

. Make sure there is at least one \tablenotemark

\tablecomments{For the AGN/QSO object classes, both secure (first number)
and unsecure (second number) identifications are counted.
}

\end{deluxetable}

\end{document}